
\let\includefigures=\iftrue

\input harvmac

\newcount\yearltd\yearltd=\year\advance\yearltd by 0

\noblackbox

\input epsf
\newcount\figno
\figno=0
\def\fig#1#2#3{
\par\begingroup\parindent=0pt\leftskip=1cm\rightskip=1cm\parindent=0pt
\baselineskip=11pt
\global\advance\figno by 1
\midinsert
\epsfxsize=#3
\centerline{\epsfbox{#2}}
\vskip 12pt
{\bf Fig.\ \the\figno: } #1\par
\endinsert\endgroup\par
}
\def\figlabel#1{\xdef#1{\the\figno}}
\def\encadremath#1{\vbox{\hrule\hbox{\vrule\kern8pt\vbox{\kern8pt
\hbox{$\displaystyle #1$}\kern8pt}
\kern8pt\vrule}\hrule}}

\noblackbox
\def\IZ{\relax\ifmmode\mathchoice
{\hbox{\cmss Z\kern-.4em Z}}{\hbox{\cmss Z\kern-.4em Z}}
{\lower.9pt\hbox{\cmsss Z\kern-.4em Z}}
{\lower1.2pt\hbox{\cmsss Z\kern-.4em Z}}\else{\cmss Z\kern-.4em
Z}\fi}
\def\IB{\relax{\rm I\kern-.18em B}}
\def\IC{{\relax\hbox{\kern.3em{\cmss I}$\kern-.4em{\rm C}$}}}
\def\ID{\relax{\rm I\kern-.18em D}}
\def\IE{\relax{\rm I\kern-.18em E}}
\def\IF{\relax{\rm I\kern-.18em F}}
\def\IG{\relax\hbox{$\inbar\kern-.3em{\rm G}$}}
\def\IGa{\relax\hbox{${\rm I}\kern-.18em\Gamma$}}
\def\IH{\relax{\rm I\kern-.18em H}}
\def\II{\relax{\rm I\kern-.18em I}}
\def\IK{\relax{\rm I\kern-.18em K}}
\def\IP{\relax{\rm I\kern-.18em P}}

\def\p{\partial}

\font\cmss=cmss10 \font\cmsss=cmss10 at 7pt
\def\IR{\relax{\rm I\kern-.18em R}}

\def\frac#1#2{{#1 \over #2}}

\def\OL#1{ \kern1pt\overline{\kern-1pt#1
   \kern-1pt}\kern1pt }
\def\vp{\varphi}
\def\[{\left [}
\def\({\left (}
\def\]{\right ]}
\def\){\right )}

\Title{\vbox{\baselineskip12pt\hbox{hep-th/0204158}
  \hbox{SLAC-PUB-9203}
\hbox{WIS/16/02-APR-DPP}\hbox{SU-ITP 02-12} }} {\vbox{ \vskip -5cm {\centerline{ Clean
Time-Dependent String Backgrounds}
\bigskip
\centerline{
 from Bubble Baths}} } }

\vskip  -5mm
 \centerline{Ofer Aharony$^{\clubsuit}$, Michal Fabinger$^{\spadesuit}$,
Gary T. Horowitz$^\diamondsuit$, and Eva
Silverstein$^{\heartsuit}$}

\bigskip
{\sl
\centerline{$^\clubsuit$Department of Particle Physics} \centerline{The
Weizmann Institute of Science, Rehovot 76100, Israel}
\medskip
 \centerline{$^{\spadesuit\heartsuit}$Department of Physics and SLAC}
\centerline{ Stanford University, Stanford, CA 94305/94309, USA}
\medskip
\centerline{$^{\diamondsuit}$Department of Physics} \centerline{
University of California, Santa Barbara, CA 93106, USA}
\bigskip
\bigskip \medskip
}


\leftskip 8mm  \rightskip 8mm \baselineskip14pt \noindent
{\tenbf \hskip 7mm Abstract.}

We consider the set of
controlled time-dependent backgrounds
of general relativity and string theory describing
``bubbles of nothing'', obtained via double analytic continuation
of black hole solutions. We analyze their quantum stability,
uncover some novel features of their dynamics, identify their
causal structure and observables,
and compute their particle production spectrum.  We
present a general relation between squeezed states,
such as those arising in cosmological particle creation, and
nonlocal theories on the string worldsheet.  The bubble
backgrounds have various aspects in common with de Sitter
space, Rindler space, and moving mirror systems, but
constitute controlled solutions of general relativity
and string theory with no external forces.
They provide a useful theoretical laboratory for
studying issues of
observables in systems with cosmological horizons,
particle creation, and time-dependent string perturbation theory.

\leftskip 0mm  \rightskip 0mm
 \Date{\hskip 8mm April 2002}

\lref\christ{D. Christodolou and S. Klainerman, {\it
The global nonlinear stability
of the Minkowski space}, Princeton University Press (1993).}

\lref\wald{R. Wald, {\it General Relativity}, University of Chicago Press
(1984) 491p.}

\lref\vishveshwara{
C.~V.~Vishveshwara,
``Stability Of The Schwarzschild Metric,''
Phys.\ Rev.\ D {\bf 1}, 2870 (1970).
}

\lref\CallanHS{
C.~G.~Callan, R.~C.~Myers and M.~J.~Perry,
``Black Holes In String Theory,''
Nucl.\ Phys.\ B {\bf 311}, 673 (1989).
}

\lref\reggewheeler{
T.~Regge and J.~A.~Wheeler, ``Stability of a Schwarzschild
Singularity'', Phys.\ Rev.\ {\bf 108}, 1063 (1957).
}

\lref\horfab{
M.~Fabinger and P.~Ho\v rava,
``Casimir effect between world-branes in heterotic M-theory,''
Nucl.\ Phys.\ B {\bf 580}, 243 (2000)
[arXiv:hep-th/0002073].
}

\lref\dggh{
F.~Dowker, J.~P.~Gauntlett, G.~W.~Gibbons and G.~T.~Horowitz,
``Nucleation of $P$-Branes and Fundamental Strings,''
Phys.\ Rev.\ D {\bf 53}, 7115 (1996)
[arXiv:hep-th/9512154].
}

\lref\BD{
N.~D.~Birrell and P.~C.~Davies,
{\it Quantum Fields In Curved Space},
Cambridge, Uk: Univ. Pr. (1982) 340p.
}
\lref\vijay{
V.~Balasubramanian, S.~F.~Hassan, E.~Keski-Vakkuri and A.~Naqvi,
``A space-time orbifold: A toy model for a cosmological singularity,''
arXiv:hep-th/0202187.
}
\lref\spacebranes{
M.~Gutperle and A.~Strominger,
``Spacelike branes,''
arXiv:hep-th/0202210.
}
\lref\BuchelWF{
A.~Buchel,
``Gauge / gravity correspondence in accelerating universe,''
arXiv:hep-th/0203041; and to appear.
}
\lref\berglundminic{
P.~Berglund, T.~Hubsch and D.~Minic,
``de Sitter spacetimes from warped compactifications of IIB string  theory,''
arXiv:hep-th/0112079.
}

\lref\noncritdS{
E.~Silverstein,
``(A)dS backgrounds from asymmetric orientifolds,''
arXiv:hep-th/0106209;
A.~Maloney, E.~Silverstein, and A.~Strominger,
work in progress.}

\lref\garyetal{
F.~Dowker, J.~P.~Gauntlett, G.~W.~Gibbons and G.~T.~Horowitz,
``The Decay of magnetic fields in Kaluza-Klein theory,''
Phys.\ Rev.\ D {\bf 52}, 6929 (1995)
[arXiv:hep-th/9507143].
}

\lref\wittenbubble{
E.~Witten,
``Instability Of The Kaluza-Klein Vacuum,''
Nucl.\ Phys.\ B {\bf 195}, 481 (1982).
}

\lref\aps{
A.~Adams, J.~Polchinski and E.~Silverstein,
``Don't panic! Closed string tachyons in ALE space-times,''
JHEP {\bf 0110}, 029 (2001)
[arXiv:hep-th/0108075].
}

\lref\twistcircles{
J.~G.~Russo and A.~A.~Tseytlin,
``Magnetic flux tube models in superstring theory,''
Nucl.\ Phys.\ B {\bf 461}, 131 (1996)
[arXiv:hep-th/9508068];
%
M.~S.~Costa and M.~Gutperle,
``The Kaluza-Klein Melvin solution in M-theory,''
JHEP {\bf 0103}, 027 (2001)
[arXiv:hep-th/0012072];
%
M.~Gutperle and A.~Strominger,
``Fluxbranes in string theory,''
JHEP {\bf 0106}, 035 (2001)
[arXiv:hep-th/0104136];
%
J.~G.~Russo and A.~A.~Tseytlin,
``Magnetic backgrounds and tachyonic instabilities in closed superstring  theory and M-theory,''
Nucl.\ Phys.\ B {\bf 611}, 93 (2001)
[arXiv:hep-th/0104238];
%
L.~Motl,
``Melvin matrix models,''
arXiv:hep-th/0107002;
%
A.~A.~Tseytlin,
``Magnetic backgrounds and tachyonic instabilities in closed string  theory,''
arXiv:hep-th/0108140;
%
J.~R.~David, M.~Gutperle, M.~Headrick and S.~Minwalla,
``Closed string tachyon condensation on twisted circles,''
JHEP {\bf 0202}, 041 (2002)
[arXiv:hep-th/0111212];
and references therein.}

\lref\empgut{
R.~Emparan and M.~Gutperle,
``From p-branes to fluxbranes and back,''
JHEP {\bf 0112}, 023 (2001)
[arXiv:hep-th/0111177].
}

\lref\mp{
R.~C.~Myers and M.~J.~Perry,
``Black Holes In Higher Dimensional Space-Times,''
Annals Phys.\  {\bf 172}, 304 (1986).
}
\lref\gpy{
D.~J.~Gross, M.~J.~Perry and L.~G.~Yaffe,
``Instability Of Flat Space At Finite Temperature,''
Phys.\ Rev.\ D {\bf 25}, 330 (1982).
}

\lref\CornalbaFI{
L.~Cornalba and M.~S.~Costa,
``A New Cosmological Scenario in String Theory,''
arXiv:hep-th/0203031.
}
\lref\NekrasovKF{
N.~A.~Nekrasov,
``Milne universe, tachyons, and quantum group,''
arXiv:hep-th/0203112.
}
\lref\ChenYQ{
C.~M.~Chen, D.~V.~Gal'tsov and M.~Gutperle,
``S-brane solutions in supergravity theories,''
arXiv:hep-th/0204071.
}

\lref\simon{
J.~Simon,
``The geometry of null rotation identifications,''
arXiv:hep-th/0203201.
}
\lref\kmp{
M. Kruczenski, R. C. Myers and A. W. Peet,
``Supergravity S-branes,''
arXiv:hep-th/0204144.}

\lref\NLST{
O.~Aharony, M.~Berkooz and E.~Silverstein,
``Multiple-trace operators and non-local string theories,''
JHEP {\bf 0108}, 006 (2001)
[arXiv:hep-th/0105309];
O.~Aharony, M.~Berkooz and E.~Silverstein,
``Non-local string theories on $AdS_3 \times S^3$ and stable
non-supersymmetric backgrounds,''
arXiv:hep-th/0112178.
}

\lref\rohm{
R.~Rohm,
``Spontaneous Supersymmetry Breaking In Supersymmetric String Theories,''
Nucl.\ Phys.\ B {\bf 237}, 553 (1984).
}



\lref\smatrix{
E.~Witten,
``Quantum gravity in de Sitter space,''
arXiv:hep-th/0106109;
S.~Hellerman, N.~Kaloper and L.~Susskind,
``String theory and quintessence,''
JHEP {\bf 0106}, 003 (2001)
[arXiv:hep-th/0104180];
W.~Fischler, A.~Kashani-Poor, R.~McNees and S.~Paban,
``The acceleration of the universe, a challenge for string theory,''
JHEP {\bf 0107}, 003 (2001)
[arXiv:hep-th/0104181].
}

\lref\dysonetal{
L.~Dyson, J.~Lindesay and L.~Susskind,
``Is there really a de Sitter/CFT duality,''
arXiv:hep-th/0202163.
}

\lref\Gott{J. Gott,
``Tachyon Singularity: A Spacelike Counterpart of the
Schwarzschild Black Hole,'' Nuovo Cimento {\bf 22B}, 49 (1974).
}
\lref\hawking{
S.~W.~Hawking,
``Particle Creation By Black Holes,''
Commun.\ Math.\ Phys.\  {\bf 43}, 199 (1975).
}
\lref\dSCFT{
A.~Strominger,
``The ds/CFT correspondence,''
JHEP {\bf 0110}, 034 (2001)
[arXiv:hep-th/0106113];
R.~Bousso, A.~Maloney and A.~Strominger,
``Conformal vacua and entropy in de Sitter space,''
arXiv:hep-th/0112218.
}
\lref\kutsei{
D.~Kutasov and N.~Seiberg,
``Noncritical Superstrings,''
Phys.\ Lett.\ B {\bf 251}, 67 (1990).
}

\lref\kachruliam{S. Kachru and L. McAllister, to appear.}

\lref\GregLaf{
R.~Gregory and R.~Laflamme,
``Black Strings And P-Branes Are Unstable,''
Phys.\ Rev.\ Lett.\  {\bf 70}, 2837 (1993)
[arXiv:hep-th/9301052].
}
\lref\SLC{
D.~Kutasov and D.~A.~Sahakyan,
``Comments on the thermodynamics of little string theory,''
JHEP {\bf 0102}, 021 (2001)
[arXiv:hep-th/0012258];
J.~Maldacena and H.~Ooguri,
``Strings in AdS(3) and the SL(2,R) WZW model. III: Correlation  functions,''
arXiv:hep-th/0111180.
}

\lref\HorowitzAP{
G.~T.~Horowitz and A.~R.~Steif,
``Singular String Solutions With Nonsingular Initial Data,''
Phys.\ Lett.\ B {\bf 258}, 91 (1991); G.~Veneziano,
``Pre-big bang cosmology: End of a myth?,''
CERN-TH-98-379
{\it 6th International Symposium on Particles, Strings and Cosmology :
PASCOS '98, Boston, MA, USA, 22 -27 Mar 1998: Proceedings / Ed. By P Nath.
-World Sci., Singapore}.}

\lref\singrefs{
J.~Khoury, B.~A.~Ovrut, N.~Seiberg, P.~J.~Steinhardt and N.~Turok,
``From big crunch to big bang,''
Phys.\ Rev.\ D {\bf 65}, 086007 (2002)
[arXiv:hep-th/0108187];
N.~Seiberg,
``From big crunch to big bang - is it possible?,''
arXiv:hep-th/0201039;
A.~J.~Tolley and N.~Turok,
``Quantum fields in a big crunch / big bang spacetime,''
arXiv:hep-th/0204091.
}

\lref\nappiwitten{
C.~R.~Nappi and E.~Witten,
``A Closed, expanding universe in string theory,''
Phys.\ Lett.\ B {\bf 293}, 309 (1992)
[arXiv:hep-th/9206078].
}

\lref\NLSTboundary{
E.~Witten,
``Multi-trace operators, boundary conditions, and AdS/CFT
correspondence,''
arXiv:hep-th/0112258;
M.~Berkooz, A.~Sever and A.~Shomer,
``Double-trace deformations, boundary conditions and spacetime
singularities,''
arXiv:hep-th/0112264;
W.~Muck,
``An improved correspondence formula
for AdS/CFT with multi-trace  operators,''
arXiv:hep-th/0201100;
P.~Minces,
``Multi-trace operators and the generalized AdS/CFT prescription,''
arXiv:hep-th/0201172;
A.~C.~Petkou,
``Boundary multi-trace deformations and OPEs in AdS/CFT correspondence,''
arXiv:hep-th/0201258;
A.~Sever and A.~Shomer,
``A note on multi-trace deformations and AdS/CFT,''
arXiv:hep-th/0203168.
}

\lref\sussfischler{
W.~Fischler, B.~Ratra and L.~Susskind,
``Quantum Mechanics Of Inflation,''
Nucl.\ Phys.\ B {\bf 259}, 730 (1985)
[Erratum-ibid.\ B {\bf 268}, 747 (1985)].
}

\lref\juanclaim{J. Maldacena, private communication.}

\lref\dalsi{
%
D.~Amati and C.~Klimcik,
``Nonperturbative Computation Of The Weyl Anomaly For A Class Of
Nontrivial Backgrounds,''
Phys.\ Lett.\ B {\bf 219}, 443 (1989);
%
G.~T.~Horowitz and A.~R.~Steif,
``Space-Time Singularities In String Theory,''
Phys.\ Rev.\ Lett.\  {\bf 64}, 260 (1990);
%
H.~J.~de Vega and N.~Sanchez,
``Space-Time Singularities In String Theory And String Propagation
Through Gravitational Shock Waves,''
Phys.\ Rev.\ Lett.\  {\bf 65}, 1517 (1990).
%
C.~R.~Nappi and E.~Witten,
Phys.\ Rev.\ Lett.\  {\bf 71}, 3751 (1993)
[arXiv:hep-th/9310112];
%
E.~Kiritsis and C.~Kounnas,
``String Propagation In Gravitational Wave Backgrounds,''
Phys.\ Lett.\ B {\bf 320}, 264 (1994)
[Addendum-ibid.\ B {\bf 325}, 536 (1994)]
[arXiv:hep-th/9310202];
%
E.~Kiritsis, C.~Kounnas and D.~Lust,
``Superstring gravitational wave backgrounds with space-time supersymmetry,''
Phys.\ Lett.\ B {\bf 331}, 321 (1994)
[arXiv:hep-th/9404114].
}

\lref\someotherbubbles{
E.g. G.~T.~Horowitz and L.~Susskind,
``Bosonic M theory,''
J.\ Math.\ Phys.\  {\bf 42}, 3152 (2001)
[arXiv:hep-th/0012037];
M.~S.~Costa and M.~Gutperle,
``The Kaluza-Klein Melvin solution in M-theory,''
JHEP {\bf 0103}, 027 (2001)
[arXiv:hep-th/0012072];
S.~P.~De Alwis and A.~T.~Flournoy,
``Closed string tachyons and semi-classical instabilities,''
arXiv:hep-th/0201185.
}

\newsec{Introduction}

Many static backgrounds are known
in which string/M theory is understood relatively
well and can be studied very concretely; some of these backgrounds are
even semi-realistic at low energies.  In particular, tori,
orbifolds, Calabi-Yau manifolds, and anti-de Sitter (AdS)
spacetimes provide very useful and
explicit backgrounds which have taught us much about the
phenomena and formalism in the theory.

Our knowledge of time-dependent backgrounds in quantum
gravity is far more rudimentary.  We do not have a
formulation of string
perturbation theory which applies to generic time-dependent
backgrounds, and we do not have a clear notion of what
the observables are in a general background (there being
in general no guarantee of the existence of an S-matrix)
\refs{\smatrix,\dSCFT}. Obviously, understanding string theory in
time-dependent backgrounds is necessary for applying string theory
to cosmology, and may be useful for relating string theory to nature.
In particular, one very
basic
process we would like to get a handle on is that of particle creation.

In order to attack these questions, we need
good examples of time-dependent solutions that
are under complete theoretical control.  Finding such
backgrounds
is more difficult than in the static case.
Time-dependent backgrounds rich
enough to exhibit particle creation are
nonsupersymmetric at low energies, since time
translation (and thus the set of
transformations generated by supercharges that commute to
time translation) is not a symmetry\foot{It would
be interesting to try to exploit the spatial
version of supersymmetry introduced in \kutsei\
to try to see if that provides any protection
from quantum instabilities.}.
A generic homogeneous cosmology
evolves from a singularity, at which point the classical spacetime
description breaks down, or evolves toward a singularity in the
future.
Aside from de Sitter space\foot{A
noncritical string construction for de Sitter space is under investigation
\noncritdS.}, the standard examples used for studying the
behavior of quantum field theory in time-dependent curved
spacetimes \BD\ are either singular (such as
cosmological and black hole geometries), involve external forces
put in by hand (such as in Rindler space or moving mirror
examples), or simply are not solutions to the Einstein equations
even at long distances.  String/M theory may ultimately
resolve the singularities in the first set of cases,
but at present we do not even know how to formulate string theory
in any of these cases. It seems reasonable to start by
considering nonsingular time-dependent backgrounds.
Although general theorems \christ\ guarantee a large class of such
solutions, e.g., describing the nonlinear scattering of
gravitational waves, essentially none are known explicitly.

In this paper we study a relatively simple set of time-dependent
non-singular solutions to general relativity and string/M theory
which exhibit a number of interesting phenomena. We start with the
observation \garyetal\ that double analytical continuations of rotating,
uncharged black hole solutions \mp, which have been much
studied as part of destabilizing instanton processes following
\wittenbubble, constitute a set of interesting time-dependent
backgrounds in their own right, which can be stabilized against
quantum mechanically induced tadpoles.  These spacetimes are
smooth and geodesically complete, and
they look like ``bubbles of nothing'' in an asymptotically flat space,
which contract until they reach a finite (tunably large in string units)
radius, at which point they start to expand.  These geometries have
aspects
in common with various features of de Sitter space, Rindler
space, and moving mirror configurations, but they constitute weakly
curved solutions of general relativity and string/M theory (which can
be arbitrarily weakly coupled in the string theory case).

These solutions have a compact (periodic) dimension, with fermions
obeying non-supersymmetric boundary conditions as they go around it, and
we find  an interesting connection between the dynamics of the bubble
and the asymptotic behavior of the compact dimension (far from the
bubble). When
the compact dimension remains finite, the bubbles continue to
expand exponentially indefinitely.
However, when the compact dimension opens up (and the asymptotic space
looks locally like Minkowski space),
the bubbles slow down and stop accelerating.
This connection is found
for the
solutions in all dimensions. When these solutions
are embedded in supergravity or string theory, this is equivalent to
saying that
the bubbles expand exponentially into regions with broken supersymmetry,
but ultimately stop accelerating in directions in which there
is asymptotic local supersymmetry (namely, the supersymmetry breaking
effects go to zero asymptotically).

For the analytic continuation of  the Schwarzschild black hole
\wittenbubble \foot{For some further discussion of various bubble solutions
see \someotherbubbles.}, the size of the compact dimension remains finite
asymptotically, and
the bubble continues to expand exponentially. We will show that this
solution has cosmological horizons (similar to de Sitter space) of
infinite
area (in contrast to de Sitter space). We will argue that this spacetime
is classically stable (at least in four dimensions). Quantum
mechanically, it is known to be unstable \rohm. So, although it provides an
interesting example of a stable classical solution to string theory with
cosmological horizons\foot{The curvature can be made arbitrarily small, in
which case
the $\alpha'$ corrections should only shift the solution negligibly
given the absence of tachyonic or marginal modes.}, it is ultimately not
a suitable background.

On the other hand, for analytic continuations
of even dimensional
Kerr black holes with all rotation parameters nonzero, the compact
dimension opens up in all directions. The
 geometry  near the bubble begins (and ends) in a phase similar
to a Milne universe, with spatial directions
near the bubble contracting (expanding) linearly in
proper time, and
the region far from the bubble reducing to flat Minkowski space.  In
between
these periods of mild contraction and expansion,  for a tunably
long period, the bubble geometry contracts and then expands exponentially.
This solution has no horizons and we will argue that perturbative
quantum corrections
do not destabilize it (assuming it is stable classically).
The S-matrix in these backgrounds is well defined.

Perhaps of most interest are the intermediate cases of bubbles obtained
from even dimensional Kerr black holes which rotate in some directions but
not in others (obviously, this requires at least six dimensions). These
are hybrid examples in which the bubble accelerates eternally in some
directions, but stops in other directions.
In particular, bubbles in
orbifolds of the ``twisted circle'' form $[(\IC^q\times
S^1)/\IZ_N]\times \IR^{8-2q, 1}$ (see \twistcircles)
experience accelerated expansion along
the $\IR^{8-2q, 1}$ directions, but remain near the origin of the complex
planes in the $(\IC^q\times S^1)/\IZ_N$ directions. We will argue that
these backgrounds are also quantum mechanically stable perturbatively,
yet have  horizons
for certain observers. These horizons are somewhat unusual since
although two such observers lose causal contact at late times, they
can both send signals to a third observer located in a different
direction. These hybrid examples are unstable to nonperturbative
quantum processes corresponding to nucleation of additional bubbles.
This presumably does not occur in the case of bubbles with all rotation
parameters nonzero, or at least, it would be very unlikely.

In addition to understanding the basic observables such as the S-matrix,
one would like to understand how to perform perturbative
field theory and string theory computations in these backgrounds.
One basic phenomenon
that arises in generic time-dependent backgrounds is particle
creation.   In quantum field theory in curved spacetime,
this involves the possibility of different choices
of vacuum for the fields.  One basic question is
how this ambiguity arises in string theory.
In section 5 we show that in general, calculating matrix elements
between squeezed states, such as those corresponding to baths of
particles generated by cosmological particle
creation, corresponds to working with a Nonlocal String Theory
\NLST\ on the string worldsheet. The bubble backgrounds provide a
textbook example of particle creation, with early and late epochs
of mild time-dependence (and Minkowski null infinity) interrupted
by a long epoch of stronger time-dependence. We compute particle
creation in field theory for modes of wavelength shorter than the
size of the bubble, and translate this to the leading order in
$\alpha'$ nonlocal action on the string worldsheet.  We also
comment on a number of basic issues involved in formulating string
perturbation theory in time-dependent backgrounds.

Although we are able to address the evident quantum
mechanically induced instabilities in these (non-supersymmetric)
backgrounds, we are
not able to rule out the existence of classical tachyonic
instabilities arising from fluctuations of the metric in the
most general cases.  This is related to the fact that due to the
complication of the coupled linear perturbation equations, the
classical stability of Kerr black holes in higher than four
dimensions has not yet been established.  However, we do rule
out classical tachyons from modes of scalar fields and from
metric fluctuations in the case of the four-dimensional Schwarzschild
bubble. It would be interesting to investigate this issue further.

It would also be interesting to investigate D-branes in our backgrounds,
and to analyze double analytic continuations
of more general black holes, such as asymptotically AdS or dS black holes.

Recently, a number of other interesting time-dependent backgrounds
have been introduced and studied, for example
\refs{\berglundminic,\aps,\noncritdS,\vijay,\spacebranes,
\CornalbaFI,\BuchelWF,\NekrasovKF,\simon,
\ChenYQ,\kmp,\kachruliam}.
The
backgrounds we study here are complimentary to these in many ways;
in particular the particle creation per mode in our backgrounds
is finite as opposed
to the situation in \vijay.

A brief outline of this paper is the following. In the next section we
study the bubbles obtained by analytic continuation of the Schwarzschild
black hole. We discuss their dynamics, horizons, classical stability,
and quantum instability. In section 3, we investigate the Kerr bubbles,
starting with the simplest four dimensional case. We then move on to the
higher dimensional bubbles, discussing both those coming from fully rotating and
partially rotating black holes.
Finally,
we discuss the quantum stability of these solutions.
In section 4 we give a quantum field theory calculation of particle
creation
in the four dimensional Kerr bubble. The final section contains a
discussion
of particle creation in string theory, and of general issues associated with
defining string theory in these bubble backgrounds and more general
time-dependent backgrounds.

\newsec{Schwarzschild Bubbles}

In this section we study spacetimes obtained from double analytic
continuation of Schwarzschild black holes \wittenbubble. We will show
explicitly
that these vacuum solutions have horizons analogous to de Sitter spacetime
(as was suggested independently by Petr Ho\v rava). We also argue that
they are
classically stable. However, they do not represent good backgrounds for
string theory since there is a nonzero quantum correction to the stress
energy tensor asymptotically (this is essentially a Casimir energy) which
destabilizes the spacetime. Nevertheless, these solutions are a
convenient starting point since they are simpler and illustrate some of
the features we will see in the stable examples discussed in the next
section.

\subsec{Classical solutions}

Consider a Schwarzschild black hole in $D$ spacetime dimensions, with metric
\eqn\schwarzmet{ds^2=-\[1-\(r_0\over r\)^{D-3}\]dt^2+
 \[1-\(r_0\over r\)^{D-3}\]^{-1}dr^2
+r^2(d\theta^2+\sin^2\theta \ d\Omega^2_{D-3}),}
where $d\Omega^2_{D-3}$ is the metric on a unit $S^{D-3}$. One can
obtain a Lorentzian vacuum solution to Einstein's equations
\refs{\Gott,\wittenbubble} by the
following double analytic continuation:
\eqn\wittcont{t\equiv i\chi, ~~~~\theta-{\pi\over 2}\equiv i\tau.}
The metric becomes
\eqn\wittenmet{ds^2=\[1-\(r_0\over r\)^{D-3}\]d\chi^2+
      \[1-\(r_0\over r\)^{D-3}\]^{-1}dr^2
+r^2(-d\tau^2+\cosh^2\tau\  d\Omega^2_{D-3}).}
The radial variable is now restricted to the range $r\ge r_0$, and
regularity at $r=r_0$ requires that $\chi$ be periodic with period
$4\pi r_0/(D-3)$ (as in the standard Euclidean black hole background,
where we only make the first continuation in \wittcont).
Thus, the spacetime asymptotically has one direction
compactified on a circle.
The solution is invariant under the large symmetry group:
$U(1)\times SO(D-2,1)$.
The maximum curvature is of order $1/r_0^2$, which can
be made arbitrarily small by taking $r_0$ large. There are no
singularities
and the spacetime is geodesically complete.

\fig{A schematic depiction of the geometry at
a fixed time, with the $\chi$ circle replaced
by two points, and the $r$ and $\phi$ directions
manifest  (drawing courtesy of Petr Ho\v rava).}{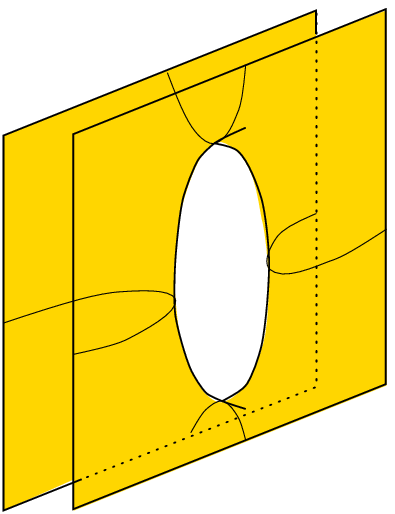}{2.3truein}

This solution describes a contracting and
then expanding ``bubble of nothing" in the following sense. Consider the
geometry on the $\tau=0$ surface. This resembles $\IR^{D-2}$ with a
sphere of radius $r_0$ removed. Over each point is a circle whose radius
smoothly goes to zero at $r=r_0$ (see Fig. 1).
Thus $r=r_0$ is not a boundary of the
space, but it is the $S^{D-3}$ of minimal area. As we move away from the
$\tau=0$ surface (both to the future and past) the area of this minimal
sphere
grows exponentially. We will call this minimal surface, given by
$r=r_0$, the ``bubble". The event horizon of the original
black hole becomes the bubble after the analytic continuations.
Note that the geometry traced out by the bubble is precisely de Sitter
space.
In effect, this solution embeds de Sitter space into an asymptotically
flat
vacuum solution.
Even though the bubble appears to be undergoing
constant acceleration, the curves at $r=r_0$ (and constant point on
$S^{D-3}$)
are geodesics. This follows from the fact that $r=r_0$ is the fixed point
of
the $\p/\p \chi$ symmetry. If these curves had a nonzero acceleration, it
would be a preferred vector orthogonal to $\p/\p \tau$ at $r=r_0$,
and there are no such
preferred vectors.

\subsec{Horizons}

We now describe the causal structure of the Schwarzschild bubble
\wittenmet.
We will show that this spacetime has (observer dependent)
horizons analogous to
de Sitter spacetime. More precisely, we show that any two observers which
stay at different points on $S^{D-3}$ lose causal contact at late time
(they do not have to stay on the bubble $r=r_0$).
It clearly suffices to consider  null curves from one observer to the
other
which move in the ($r,\chi,\vp$) directions where $\vp$ is a parameter
along
the geodesic in $S^{D-3}$ connecting the position of the two observers.
Given such a  curve $x^\mu(\lambda)$ with tangent vector $\xi^\mu = \dot
x^\mu$, the condition that it be null, $\xi^\mu \xi_\mu =0$, is
\eqn\nullcond
{ \dot \chi^2 f(r) + {\dot r^2\over f(r)}
- \dot \tau^2 r^2 + \dot \vp^2 r^2 \cosh^2 \tau =0,  }
where
\eqn\deff{f(r) \equiv 1 - \({r_0\over r}\)^{D-3}.  }
It follows that
\eqn\ineqvp{
\dot \vp^2  \cosh^2 \tau \le \dot \tau^2.}
If the null curve starts at some late time $\tau_0$, we have
$|\dot \vp| \le e^{-\tau}
\dot \tau$, so the maximum change in $\vp$ to the future is
$|\Delta \vp_{\rm max}| = e^{-\tau_0}$.
This shows that  any two observers that stay at different $\vp$ will lose
causal contact at late times.

This result is easy to understand intuitively when both observers are
sitting on the bubble $r=r_0$.
We know that de Sitter space
has horizons, so two observers that stay at different points on the
$S^{D-3}$
cannot communicate by null curves that stay on the bubble. But a
null curve that moves in the $r,\chi$ plane projects onto a timelike curve
in the de Sitter space, and hence has even less chance of being used to
send
a signal between observers at different points on the sphere. Of course it
would be very strange if one could send a signal by a null curve but not
by a
null geodesic. This does not happen. Within the
de Sitter space at $r=r_0$, the boundary of the region that
can communicate with an event p consists of (de Sitter)
null geodesics from p. One can show
that these geodesics are also null geodesics of the full bubble spacetime.
In general, null geodesics which start at constant $r$ and $\chi$, stay
at constant $r$ and $\chi$.

We now show that the only restriction on whether observers can
communicate at late time is the one we have just discussed:
that they have the same coordinates on the $S^{D-3}$.
 Let $\xi^\mu$ now be tangent to a null geodesic
moving in the $r,\chi$ plane, but not moving on the $S^{D-3}$. There are
two
conserved quantities $P_\chi = \dot \chi f(r)$ and
$E= r^2 \dot \tau$ \foot{Even
though $\p/\p \tau$ is not a symmetry of \wittenmet,
it is a symmetry of the three dimensional spacetime obtained by fixing a
point on $S^{D-3}$.}. So the condition $\xi^\mu \xi_\mu =0$ now yields
\eqn\geod{
\dot r^2 -{E^2\over r^2}f(r) = - P_\chi^2.  }
The second term is an effective potential which vanishes both at $r=r_0$
and at $r=\infty$. So, typical null geodesics oscillate between a maximum
and minimum value of $r$. Clearly, by changing $E$ and $P_\chi$, one can
find null geodesics which connect observers at any two values of $r$.
Similarly, from the definition of $P_\chi$ we see that $|\dot \chi |\geq
|P_\chi|/ f(r_{max})$. So, as long as $P_\chi \ne 0$, the null geodesic
goes around the $\chi$ circle infinitely many times and can easily connect
observers at different $\chi$.

What is the Penrose diagram of the Schwarzschild bubble? This is a little
subtle. In drawing Penrose diagrams, one usually suppresses the spheres
of spherical symmetry. However, if we do this, we will not see the
horizons, since they only exist for observers at different points on the
sphere. So we will keep one direction on the sphere. Another problem is
that due to the Kaluza-Klein boundary conditions, one cannot conformally
rescale the metric and add a smooth boundary at null infinity. We will
avoid this by suppressing the $\chi$ direction. The remaining spacetime is
shown in Fig. 2.
\fig{Penrose diagram of Schwarzschild bubble.  The
space inside the shaded region is absent; its surface at $r=r_0$ is
the bubble which traces out de Sitter space.}{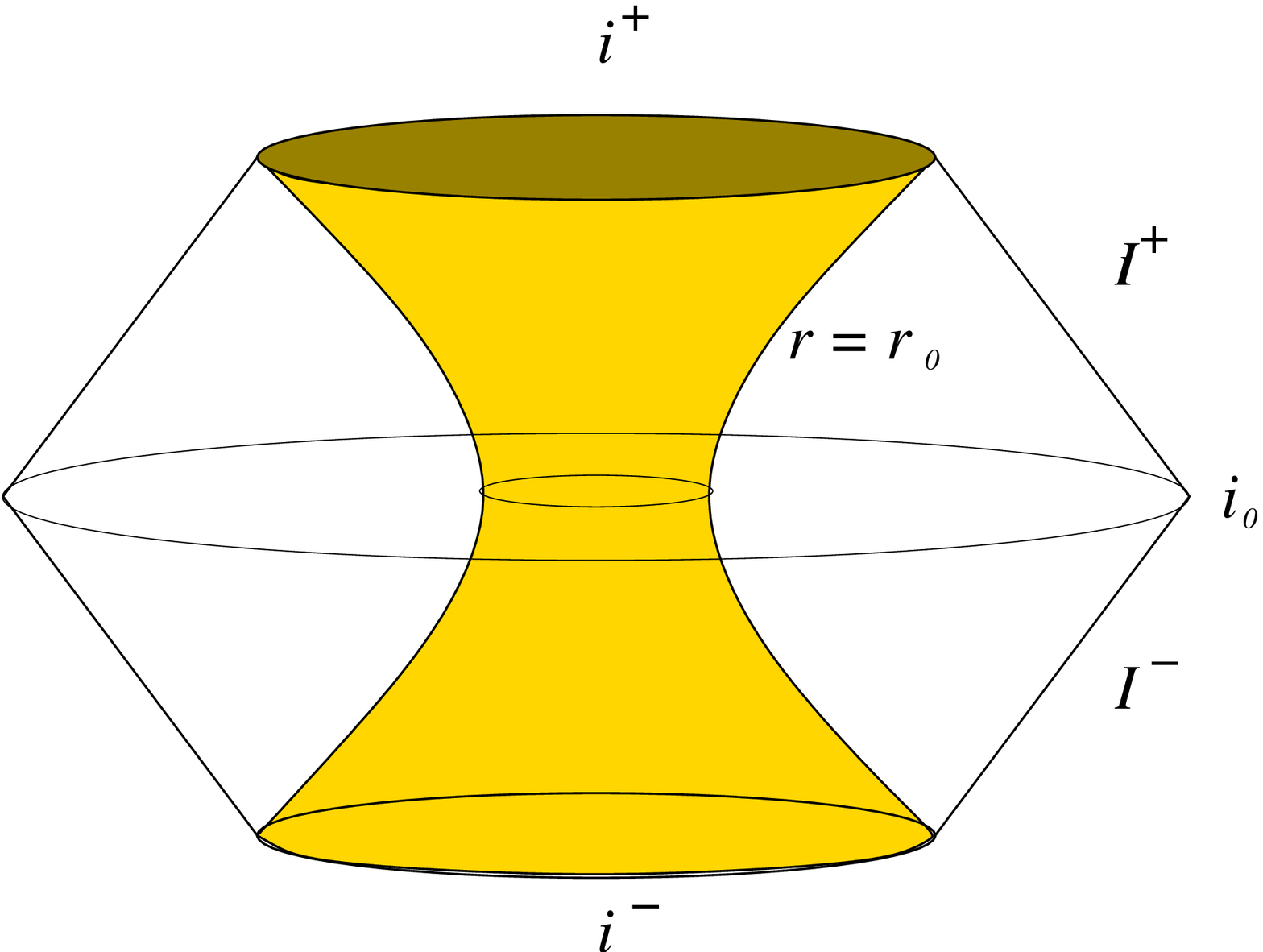}{4truein}
 It is conformally equivalent to the region
$X^\mu X_\mu \ge 1$ in Minkowski spacetime. One can think of this as a
spacetime in which future and past timelike infinity, $i^+, i^-$, are no
longer points, but spread out into spacelike circles (or spheres in the
full spacetime). It is easy to verify that timelike geodesics also
oscillate between maximum and minimum values of $r$, so all timelike
geodesics
have endpoints on $i^-$ and $i^+$.

The Schwarzschild bubble has no event horizons in the black hole sense:
the past of future null infinity includes the entire spacetime. This is
clear since every event can send a light ray to at least some points on
${\cal I}^+$, e.g. the generator with the same position on $S^{D-3}$. (But
unlike Minkowski spacetime, the light rays from an event p do not reach an
entire cross-section of ${\cal I}^+$. Some of the null generators of
${\cal I}^+$ cannot receive signals from p.) However, as we have seen,
there are cosmological horizons analogous to de Sitter space in the sense
that the past of every complete timelike geodesic is not the entire
spacetime.
As in de Sitter space, if we consider the vacuum determined by
analytic continuation of the path integral on the Euclidean
black hole background, static observers on the bubble  would see
a thermal bath with a temperature
of order $1/r_0$ (so they may consider themselves to be in a bubble bath).
One important difference from de Sitter space concerns the
horizon
area. If we define the area of the horizon by considering the boundary of
the past of the endless geodesic (as one does in asymptotically de
Sitter spacetimes), it is clear that the horizon area is infinite. So,
there does not seem to be a finite dimensional Hilbert space associated
with these spacetimes, and the issue raised in \dysonetal\ does not arise.

\subsec{Classical stability}

The metric \wittenmet\ is a classical solution to Einstein's equations.
Let us now investigate its stability, beginning at the classical level.
This requires checking whether there are normalizable
effectively tachyonic modes
(namely, modes which grow in time relative to the background metric) among
the solutions to the linearized field equations. Such modes would be
localized near the bubble since the asymptotic region has
no such excitations.

We begin by considering a scalar field $\phi$.
In this background, it satisfies the following equation of
motion (where we set $r_0=1$ for simplicity, it can always be restored
by dimensional analysis) :
\eqn\eom{ {1\over{(1-{1\over r^{D-3}})}}
\del_{\chi}^2\phi + {1\over r^{D-2}}\del_r[r^{D-2}(1-{1\over
r^{D-3}})\del_r\phi] +{1\over r^2}\nabla^2_{dS_{D-2}}\phi = m^2\phi, }
where $m^2$ is the bare mass.  Let us separate variables and consider
modes
with definite de Sitter mass $M^2$,  $\nabla^2_{dS_{D-2}}\phi = M^2\phi$,
and definite momentum $k$ around the $\chi$ direction,
$\del_\chi^2\phi=-k^2\phi$.  For now let us just consider modes with $k=0$
and particles with zero mass, $m=0$.

The solutions at large and small $(r-1)$ are as follows.  At
large $r$, we have solutions
\eqn\larger{
\phi\sim r^{-{(D-3)\over 2}\pm i\mu},
}
where as in \dSCFT, $\mu=\sqrt{M^2-{(D-3)^2\over 4}}$.
The solution which goes as $\phi \sim
r^{-(D-3)/2-|\mu|}$ (for $M^2 < {(D-3)^2\over 4}$)
becomes more and
more normalizable as $M^2$ becomes more negative, and it
is normalizable for $M^2 < {(D-3)^2\over 4}-1$.

For $r$ near the bubble, by changing coordinates to $r=1+\rho^2$,
we find a series solution
\eqn\smallr{
\phi=\phi_0(1-{M^2\over (D-3)}\rho^2+\dots)=\phi_0(1-{M^2\over
(D-3)}(r-1)+\dots),
}
and another singular solution which diverges logarithmically.
We want to know if there is any value of $M^2$ such that the
smooth series solution \smallr\ matches onto
the normalizable solutions $\phi \sim r^{-(D-3)/2-|\mu|}$,
with $M^2< {(D-3)^2\over 4}-1$.
In fact, it is easy to see that the full solution is monotonic and
thus \smallr\  cannot match onto the decaying
solution at large $r$, for any $M^2<0$.
Let us take without loss of generality
$\phi(1)=\phi_0= 1$.
Our equation of motion \eom\ is (again for $m=k=0$)
\eqn\eomII{
{1\over r^{D-2}}\del_r[r^{D-2}(1-{1\over r^{D-3}})\del_r\phi]
=-{1\over r^2}M^2\phi,}
which can be written in the form
\eqn\eomIII{
{1\over r^{D-2}}\del_r[a(r)\del_r\phi]
= {1\over r^{D-2}}(a'\phi' + a\phi'')=-{1\over r^2}M^2\phi,
}
where $a>0$ and $a'>0$ everywhere.  Now, from \smallr\ we know that
for $M^2<0$, $\phi'$ starts out positive, so the mode begins
growing from its starting value of $\phi_0=1$ at the bubble.
The question is then whether it can turn around, reaching
some maximum positive value of $\phi$, $\phi_c\equiv\phi(r_c)$
for some $r_c$.  This would
require $\phi'(r_c)=0$.  The equation \eomIII\
would then be at $r_c$
\eqn\eomIV{
{1\over r^{D-2}}\del_r[a(r)\del_r\phi]\bigg|_{r=r_c}
= {1\over r^{D-2}}(a\phi'')\bigg|_{r=r_c}=-{1\over r_c^2}M^2\phi.
}
But, since the right-hand side of this is greater than zero for negative
$M^2$, so is the left-hand side.  Since $r>0$ and $a>0$ always,
this would mean we would have to have $\phi'' >0$,
a contradiction.  So there can be no point where the
mode turns over, and the large $r$ continuation of \smallr\ must
always include the nonnormalizable
solution at large $r$.  By a simple rescaling of $\phi$, this argument
generalizes to arbitrary $m$ ($m^2 > 0$) and $k$, and to positive
$M^2$ with $M^2 < (D-3)^2/4 -
1$.

It also appears to
generalize, given the
metric perturbations of the $D=4$ Schwarzschild black hole studied
explicitly
in \refs{\reggewheeler,\vishveshwara}, to metric perturbations
in $D=4$, at least for $k=0$ (higher $k$ modes might be expected
to be less tachyonic in any case).  One can translate the modes
studied in \refs{\reggewheeler,\vishveshwara},
expressed in terms of tensor spherical
harmonics in the black hole background, to modes in the bubble
geometry.  In the former case, the spherical harmonics are
eigenfunctions of the Laplacian on the spherical directions
of the black hole with eigenvalue $-l(l+1)$; in our analytic
continuation these become tensor spherical harmonics on the
de Sitter slices of the Schwarzschild bubble geometry with
$M^2_{dS}=-l(l+1)$.  The time direction in the black hole
becomes our $\chi$ direction, and imaginary frequency for the
black hole modes (corresponding to tachyons in that geometry)
would correspond to real momentum $k$ along our $\chi$ circle,
which for us is quantized.
The analysis of \refs{\reggewheeler,\vishveshwara} rules out tachyons
in the black hole background by showing that there
are no normalizable and nonsingular solutions to the
radial part of the equations of motion, and this
directly rules out tachyonic modes in the bubble of $M^2_{dS}=-l(l+1)$.
Furthermore, we find that by replacing $-l(l+1)$ by $M^2$ in
their analysis, and rescaling fields appropriately, one
can extend their arguments to arbitrary $M^2$, at least for
$k=0$.
Thus, at least in four dimensions, the Schwarzschild bubble
appears to be a classically stable solution to the equations of motion.

In $D>4$ dimensions, there are two possibilities. One can consider the $D$
dimensional bubble solution \wittenmet, or one can take a product of a
lower dimensional bubble and another Ricci-flat solution
such as flat space. In the latter
case there is a possible instability analogous to the Gregory-Laflamme
instability of black strings \GregLaf. This arises since the Euclidean
Schwarzschild metric has a negative mode \gpy. In other words, there is a
transverse,
traceless $h_{\mu\nu}$ satisfying $\triangle_L h_{\mu\nu} = -\lambda^2
h_{\mu\nu}$, where $\lambda^2$ is of order $1/r_0^2$ and $\triangle_L$ is
the Lichnerowicz operator obtained from the linearized Einstein equation.
This mode is spherically symmetric and independent of $\chi$. Thus, it
can be analytically continued to a real perturbation on either the
Schwarzschild black hole or the bubble spacetime. Of course, by itself, it
does
not define a physical perturbation of the Lorentzian spacetime since it
does not satisfy the linearized vacuum field equations ($\triangle_L
h_{\mu\nu} = 0$). But if there are extra flat directions with coordinates
$x^i$, one can consider $h_{\mu\nu} e^{iq_i x^i}$ with $q^2 =
\lambda^2$. For the black string, this is a static perturbation. For the
bubble, it is invariant under the de Sitter symmetry. For the black
string, it has been shown explicitly that this value of $q^2$ is the
dividing line between stable and unstable perturbations: smaller $q^2$ are
unstable while larger $q^2$ are stable \GregLaf. It is likely that the
same
will be true for the ``extended bubbles''. To avoid this instability, one
must assume the extra dimensions are compactified and have size smaller
than $r_0$,
so that every $q^2 \ne 0$ is bigger than $\lambda^2$.

\subsec{Quantum instability}

Since the $\chi$ circle at infinity is contractible, there is only one
choice of spin structure for fermions\foot{In the case of $D=4$ we
are choosing the ordinary (trivial) spin structure for the $\phi$ circle.}.
This corresponds to fermions which
are antiperiodic around the circle, and hence supersymmetry is broken
even asymptotically far away from the bubble by the
Scherk-Schwarz mechanism. In string theory this type
of supersymmetry breaking was studied by Rohm \rohm, and the same analysis
applies to our backgrounds at large $r$. For large enough $r_0$, there
are no tachyons from the string winding sector at infinity. However, at
1-loop order in string perturbation theory, a tadpole develops for the
radius of $\chi$, driving it to smaller values, toward the tachyonic
regime (in analogy to the case of a non-supersymmetric compactification of
M-theory \horfab). The case of string theory corresponds to $D=10$, and the 1-loop
induced energy density there scales as $-1/r_0^{10}$ \rohm.
In addition, there is
a nonperturbative quantum instability corresponding to the nucleation
of additional bubbles far from the one we are studying\foot{In the case of
M-theory it has been argued that this is closely related (in the regime of small $r_0$)
to the uncharged closed string tachyon instability \horfab.}. The calculation of
the rate for this
process is identical to the original calculation of the decay of the
Kaluza-Klein vacuum \wittenbubble.
It is easy to see that the perturbative instability
dominates over the nonperturbative instability.
These instabilities mean that the background \wittenmet\ does not provide
a
useful controlled time-dependent background (beyond the classical theory).
However, there are generalizations which do, to which we now turn.

\newsec{Kerr Bubbles}

One can perform a similar double analytic continuation of rotating (Kerr)
black holes, as discussed in \garyetal. Here we will study the dynamics
of these solutions, uncovering some important new features. In the
case where all angular momenta are nonzero, we argue
that this construction leads to quantum-mechanically stable time-dependent
backgrounds, assuming there are no classical tachyons from metric
perturbations\foot{The latter we are not able to check directly; this is
related
to the fact that this computation has so far proven to be prohibitively
difficult also for higher-dimensional Kerr black holes, which are also not
known
for sure to be stable.}. In the case where some (but not all)
the angular momentum parameters
are nonzero, we show that the solutions
contain horizons for
certain (but not all) observers. While these latter solutions do not appear to
have
perturbative quantum instabilities, they may still have nonperturbative
instabilities associated with the nucleation of additional bubbles.

\subsec{Four dimensional Kerr bubble}

Let us start with the simplest case of the four-dimensional
Kerr black hole.  If one takes the Kerr metric in Boyer-Linquist
coordinates
and performs the double analytic continuation \wittcont\
together with an analytic continuation of the angular momentum
parameter $a \rightarrow i\beta$, one obtains the metric
\eqn\KerrbubbleIV{\eqalign{
ds^2 = & - (r^2 + \beta^2 \sinh^2\tau) d\tau^2 + d\chi^2 +
       (r^2 - \beta^2) \cosh^2\tau d\phi^2 \cr
& -       {{r_0 r} \over {r^2 + \beta^2 \sinh^2\tau} }
                                (d\chi + \beta \cosh^2\tau d\phi)^2
     + {{r^2 + \beta^2 \sinh^2\tau}\over {r^2 - \beta^2 - r_0 r}} dr^2.\cr
}}
The radial coordinate is restricted to  $r\ge r_b$ where $r_b$ is the
larger solution to the equation
$r^2 - \beta^2 - r_0 r = 0$. This minimal radius again
describes a time-dependent bubble. In this case, one finds that regularity
at the bubble requires that we make identifications\foot{For the $\phi$
circle we will choose a trivial spin structure and period $2\pi$, although
there are other possibilities. One could go to the
universal covering space to
make $\phi$ non-compact and then, if desired, one could compactify it on a
circle of any period and  spin structure.}
by the following symmetry operator
\eqn\KerrIdOp{\hat O =  e^{ 2\pi R \ \!  \partial_\chi} e^{2 \pi  R\Omega
\ \! \partial_\phi} (-1)^F,}
where $R=2r_0 r_b/\sqrt{r_0^2 + 4\beta^2}$ is the inverse of the surface
gravity of the Kerr black hole (after the Wick rotation), $\Omega=\beta /
r_0 r_b$, and $F$ is the spacetime fermion number. This means that
we need to have coordinate identifications
 \eqn\KerrID{ (\chi,\phi)\equiv(\chi+2\pi n_1R, \phi +2\pi n_1R\Omega
+2\pi n_2), \quad n_1, n_2 \in \IZ.}
At a fixed time, this metric describes a ``bubble of nothing'' excising a
region near the origin of the orbifold of flat space
\eqn\twistline{(\IC\times \IR)/\Gamma,} where $\Gamma$ is the group of
identifications generated by \KerrIdOp\ (where $\chi$ parameterizes the
$\IR$ factor and $\phi$ the angular direction in the complex plane $\IC$
as in Fig. 3).

\fig{The orbifold of flat space corresponding to \KerrID. The identification of
the two darker planes in the figure involves a $2\pi R \Omega$ rotation of $\phi$.}
{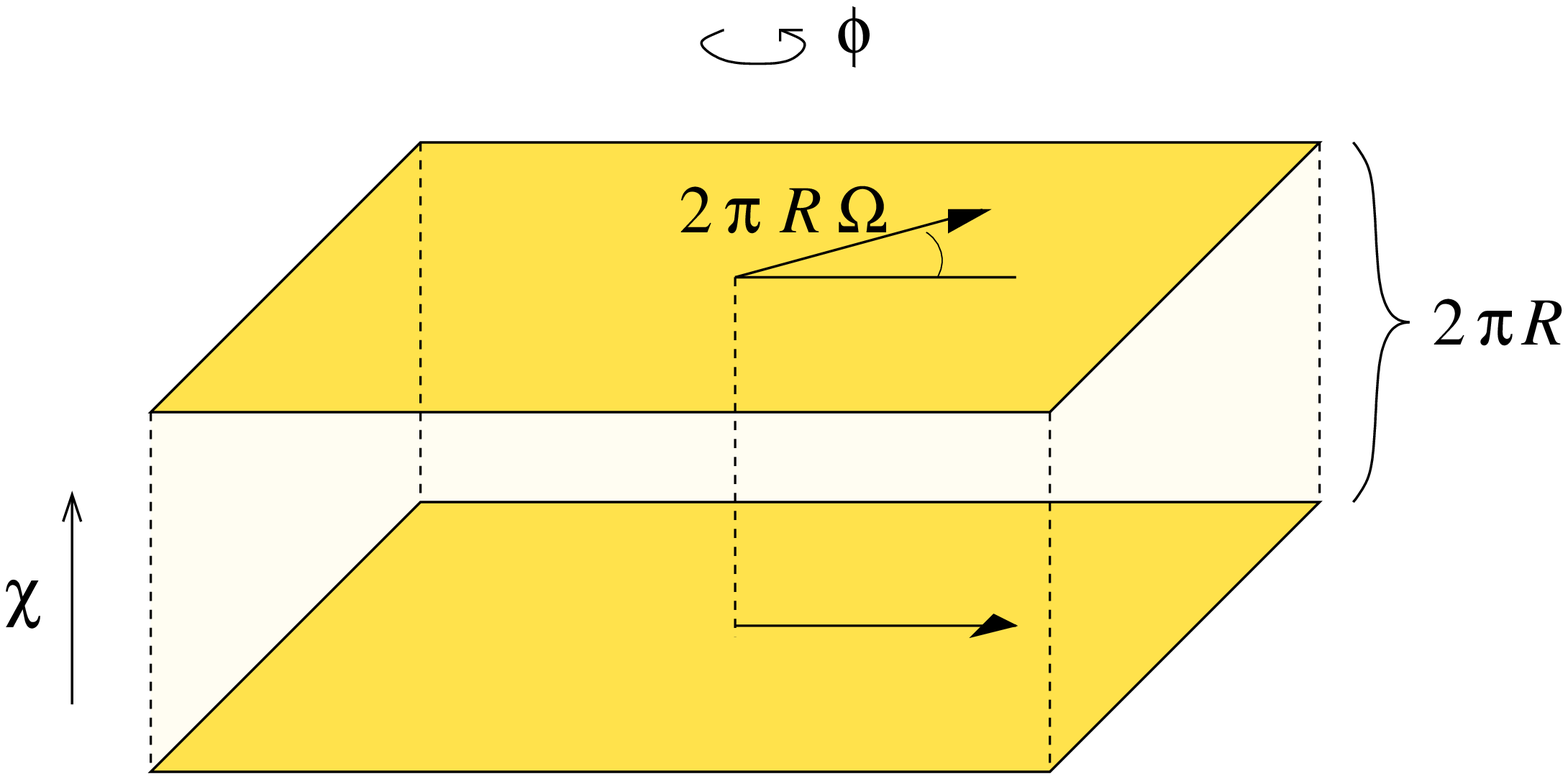}{3.5truein}

For rational values of $R \Omega$, the identifications
\KerrID\ produce a ``twisted circle'' compactification of the sort
recently studied in \twistcircles. Note that this Kerr bubble is not rotating.
There is no $d\tau d\phi$ term in the metric. The parameter $\beta$ just
describes the spatial identification.

An important difference between the identification \KerrID\ and the
simpler identification of the Schwarzschild bubble is that now the
compact dimension opens up asymptotically. In other words, the size of
the compact circle grows with $r$. This has an important implication
for supersymmetry,
namely, if we consider \KerrbubbleIV\ as a solution to supergravity then
the space is asymptotically {\it locally} supersymmetric (in
the sense of the SUSY breaking effects being suppressed at large $r$).
It is only locally supersymmetric at infinity
because a spinor transported all the way around the
circle defined by \KerrIdOp\
does not come back to itself, but the size of this circle goes to
infinity as $r \to \infty$.

Let us look at the evolution of this bubble.  For this it is useful to
study the induced metric on the bubble $r=r_b$. The coordinate
$\tilde \phi = \phi -\Omega \chi$ does not change under the identification
\KerrID, so it is a natural coordinate on the bubble, and the induced metric
is given by
\eqn\nearKerr{ ds^2_b =
-(r_0 r_b + \beta^2 \cosh^2\tau) d\tau^2 +
       {{r_0^2 r_b^2\cosh^2\tau}\over {r_0 r_b + \beta^2 \cosh^2\tau}}
                                    d\tilde \phi^2.
}
For $\beta^2 \cosh^2\tau \ll r_0r_b$, the bubble evolves
like de Sitter space, as in the case of the Schwarzschild bubble.
However, for $\tau\to\infty$ the bubble metric becomes
\eqn\latenearKerr{
ds^2_b\to -\beta^2 d(e^\tau)^2+ {{r_0^2 r_b^2 }\over\beta^2}d\tilde \phi^2.
}
This is simply a circle staying at a fixed proper radius:  the
bubble has stopped expanding!  Similarly, for $\tau\to -\infty$,
the bubble stays at a fixed size, and only begins contracting
appreciably
when $\beta^2\cosh^2\tau$ is comparable to $r_0r_b$.

The behavior of the full metric at late times can be most easily
seen by changing variables to $t=\sinh\tau$.  The metric is then
\eqn\KerrIVt{\eqalign{
ds^2 = &-{{r^2 + \beta^2 t^2}\over {1 + t^2}} dt^2 + d\chi^2 +
( r^2 - \beta^2) (1
+ t^2) d\phi^2\cr
& -{{r_0 r}\over {r^2 + \beta^2 t^2}} (d\chi + \beta (1 + t^2)
d\phi)^2 +
            {{r^2 + \beta^2 t^2}\over {r^2 - \beta^2 - r_0 r}} dr^2.\cr
}}
At late times we see from this metric that the space around the
bubble continues to expand, but with a local scale factor
linear in the proper time (similarly to the Milne universe)
rather than exponential.  The acceleration of the de Sitter-like
phase has stopped, and we
enter a phase of mild time-dependence which we will refer
to as the ``Milne'' epoch.
\fig{Penrose diagram of the four dimensional Kerr bubble, with
null rays $a$ and $b$ separating the ``Milne'' and ``de Sitter''
epochs.}{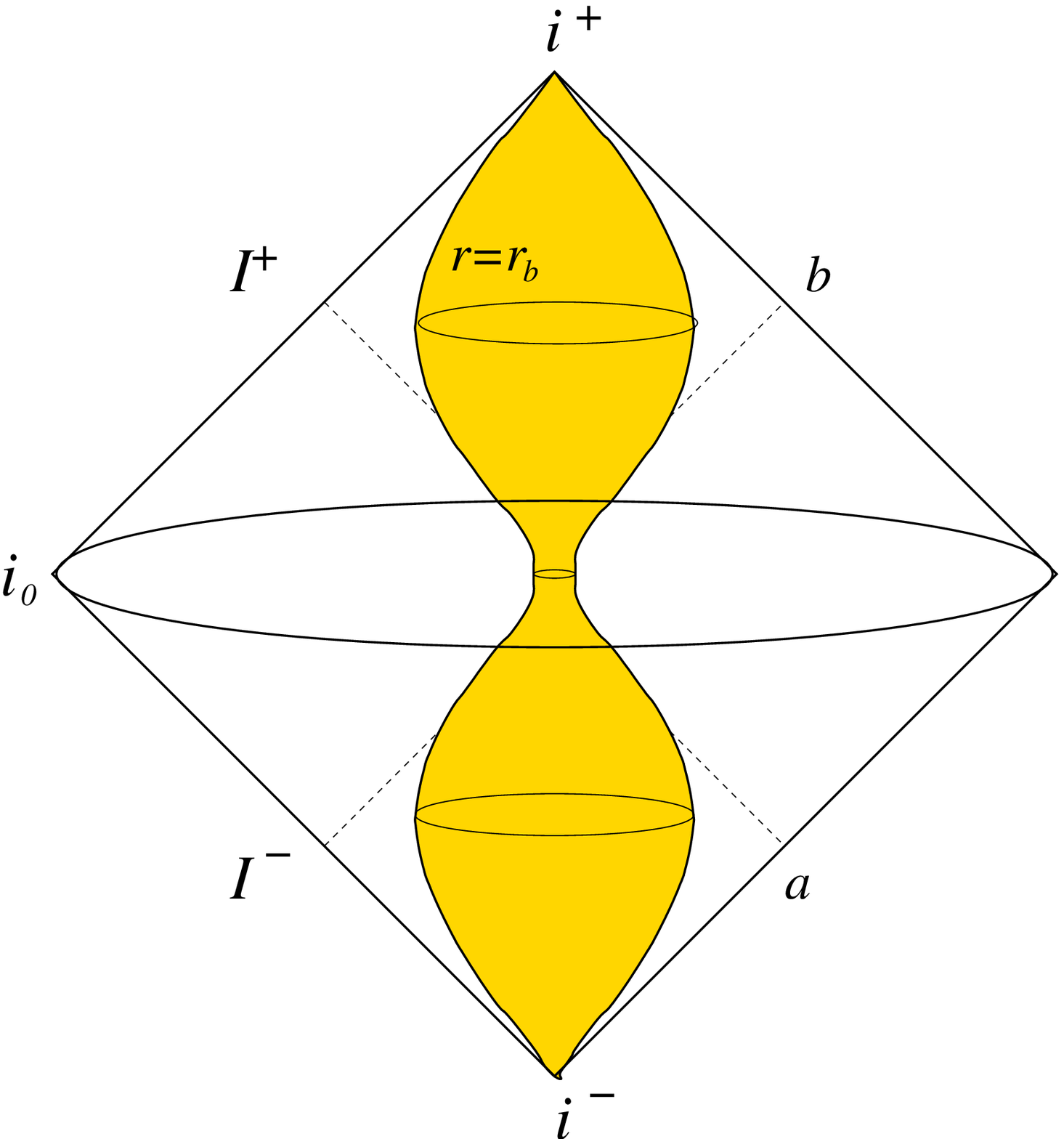}{4truein}
The Penrose diagram is shown in Fig. 4, with
the asymptotic region equivalent to Minkowski space, in particular
with a complete Minkowski null infinity. It is clear that this solution has
no horizons. However, it is
extremely well suited to studying the phenomenon of particle creation:
it is an everywhere smooth weakly coupled background with
mild time dependence in the past and future (and Minkowski
null infinity), interrupted by a phase (the de Sitter phase)
of strong time dependence.  We will compute particle creation
for some modes in this background in \S4, and interpret
particle creation in string theory in \S5.

Since the Kerr bubble behaves qualitatively differently from the
Schwarzschild
bubble, it is natural to wonder if this is related to another
qualitative difference between the spacetimes. Namely, in Kerr, the
compact dimension opens up asymptotically, while in Schwarzschild it
does not.
We can check this by going to higher dimensions where there are solutions
in which the compact dimension opens up in some directions and remains
finite in others. We will find that in all cases, the bubble
continues to accelerate in directions where the compact dimension
approaches a constant radius, but stops accelerating in directions where
the
compact dimension opens up. It is as if the bubble ``loses energy"
when the dimension opens up.

\subsec{Higher dimensional Kerr bubbles}

We now turn to the
dynamics of the higher-dimensional Kerr bubbles, which
is our main case of interest. We start with the rotating black holes
found by Myers and Perry \mp.
The form of the metric differs in
odd and even dimensions. We will consider the case of even spacetime
dimension $D$. The metric has parameters associated with rotation in
different orthogonal planes, so it is convenient to introduce spatial
coordinates
based on $D/2 -1$ orthogonal planes: ($\mu_j, \phi_j$) with $\mu_j\ge 0$
and
  $0\le \phi_j \le 2\pi$.
There is one remaining spatial
direction with coordinate $\alpha$. Rather than work with the separate
radial variables $\mu_j$ and $\alpha$, Myers and Perry introduce an
overall radial variable $r$ and impose the constraint
\eqn\constraint{
\sum_j \mu_j^2 + \alpha^2 =1,}
so that $-1 \le \alpha \le 1$ and  $0\le \mu_j \le 1$.

The even dimensional Kerr black hole is \mp\
\eqn\evenkerr{\eqalign{
ds^2 = &-dt^2 + r^2 d\alpha^2 + \sum_j(r^2 + a_j^2)(d\mu_j^2 + \mu_j^2
d\phi_j^2) \cr
&+{r_0^{D-3} r\over \tilde\Pi \tilde F}(dt +\sum_j a_j \mu_j^2 d\phi_j)^2
   + {\tilde\Pi \tilde F\over \tilde\Pi -r_0^{D-3} r} dr^2,}}
where
\eqn\defF{
 \tilde F(r,\mu_j) = 1-\sum_j {a_j^2 \mu_j^2\over r^2 + a_j^2},
   \qquad \tilde\Pi(r) = \prod_j (r^2 + a_j^2).  }
To obtain the Kerr bubble, we perform the analytic continuation
\eqn\analyticcon{ t=i\chi, \quad a_j = i\beta_j, \quad \alpha =
i\sinh\tau,}
giving
\eqn\evenkerrbub{\eqalign{ ds^2 = &- r^2\cosh^2\tau d\tau^2
+d\chi^2 + \sum_j(r^2 - \beta_j^2)(d\mu_j^2 + \mu_j^2 d\phi_j^2) \cr
&-{r_0^{D-3}
r\over \Pi F}(d\chi +\sum_j \beta_j \mu_j^2 d\phi_j)^2
   + {\Pi F\over \Pi -r_0^{D-3} r} dr^2,}}
where now
\eqn\defF{
  F(r,\mu_j) = 1+\sum_j {\beta_j^2 \mu_j^2\over r^2 - \beta_j^2},
   \qquad \Pi(r) = \prod_j (r^2 - \beta_j^2).  }
In light of the constraint \constraint, $\mu_j$ are now time dependent. It
is convenient to extract this time dependence by setting
\eqn\defmu{\mu_j = \hat x_j \cosh\tau,}
with $\sum_j \hat x_j^2 =1$. The
minimum value of $r$ is $r_b$ which is defined to be the largest solution
to $\Pi -r_0^{D-3} r =0$. We again refer to this as the bubble. Regularity
on
the bubble requires identifications similar to \KerrID,
\eqn\KerrIDhigherD{ (\chi, \phi_j) = (\chi + 2\pi R n_0, \phi_j +
2\pi R \Omega_j n_0 +2\pi n_j), }
where
\eqn\RotationPar{\Omega_j = {\beta_j \over r_b^2 - \beta_j^2 }\quad {\rm
and} \quad R = {2r_0^{D-3} r_b\over {{\partial \Pi \over \partial r
}|_{r=r_b}
-r_0^{D-3} } }, }
again with antiperiodic fermions for the $n_0$ identification.

We now describe the basic features of the Kerr bubble. First, one can
check that for $D=4$, this solution reduces to \KerrbubbleIV.
In general, the time-time component of the metric is
\eqn\timetime{
g_{\tau\tau} = -[r^2 + \sum_j (\beta_j \hat x_j)^2\ \sinh^2\tau].}
We see the same behavior as in four dimensions. For small $\tau$, the
proper
time along curves of constant $r$ and $\hat x_j$ is proportional to $\tau$,
while for
large $\tau$, the proper time is proportional to $e^\tau$. The only
exception is if some of the $\beta_j$ vanish. Then, in the directions
orthogonal to all the planes of rotation, the second term in \timetime\
vanishes and the proper time remains proportional to $\tau$ even at late
time. It is clear that the metric components grow no faster than
$e^{2\tau}$ at late time. Thus, generically distances expand linearly with
proper
time and the spacetime is again in a Milne phase at late time. However, in
directions perpendicular to the planes of rotation, the expansion
remains exponential for all time.

In four dimensions we saw that the bubble stops expanding completely at
late
time. This is not true in higher dimensions.
One can see this by looking at
the component of the metric in the $\phi_j$ direction.
On the bubble ($\Pi = r_0^{D-3} r_b$), this is
\eqn\phiphi{
g_{\phi_j\phi_j} = {(r^2_b-\beta_j^2)\hat x_j^2 \cosh^2\tau\over F}
\[1+\sum_{k\ne j} {\beta_k^2 \hat x_k^2\cosh^2\tau \over r^2_b -
\beta_k^2}\]. }
The coefficient in front of the brackets is independent of
$\tau$ at late time. In $D=4$, there is only one rotation plane, so the
sum inside the bracket is absent. Hence $g_{\phi_j\phi_j}$ approaches a
constant and since the bubble is just this circle, the bubble stops
expanding. In higher dimensions, $g_{\phi_j\phi_j} \sim e^{2\tau}$, so it
generically expands exponentially with proper time initially, and later
slows down and expands only linearly. However if some of the rotation
parameters are zero, one can set $\hat x_j=0$ for every nonzero $\beta_j$.
This corresponds to looking in a direction orthogonal to all the planes of
rotation. In this case $F=1$, and the sum in \phiphi\ vanishes. Hence
$g_{\phi_j\phi_j}\sim e^{2\tau}$ at late time, and since $\tau$ is
proportional to proper time, the bubble continues to expand exponentially
in
these directions. Since the bubble expands exponentially in some
directions and only linearly in others, it becomes highly
distorted\foot{This phenomenon
may also occur in gravity duals to gauge theories in metastable vacua
\juanclaim.}.

The significance of the continued exponential expansion (even if it occurs
only in certain directions) is that observers in these directions will
lose causal contact and there will be (observer dependent) horizons.
To see this, we first show that the metric orthogonal to the
planes of rotation is very similar to the Schwarzschild bubble  discussed
in the previous section. Setting $\hat x_j =0$ for every nonzero
$\beta_j$, we see that the $d\chi d\phi_j$ cross terms vanish. Since
$F=1$, the resulting metric takes the form
\eqn\almostSch{ ds^2 = \(1-{r_0^{D-3}
r\over \Pi}\) d\chi^2 + \(1-{r_0^{D-3} r\over \Pi}\)^{-1}dr^2
+r^2(-d\tau^2 +
\cosh^2\tau d\Omega^2).  }
This metric differs from \wittenmet\ only in
the radial dependence of the $\chi,r$ plane. The analysis of the horizons
in section 2.2 carries over exactly with the redefinition $f(r) =
1-(r_0^{D-3}
r/\Pi)$. This shows that observers in the space orthogonal to the planes
of
rotation cannot communicate by sending signals in this space.

To see if there are horizons, we must investigate the motion of
light rays off this subspace. When one does this, one finds that at late
times
two observers can both send signals to a  third observer living off this
subspace, but the third observer cannot send signals back to them (and
hence
they cannot communicate with each other). This is illustrated by looking
at the asymptotic form of the solution near null infinity.
If we take $r\gg r_0,\beta_j$, and $\tau\gg 1$, the metric becomes
\eqn\asymbubble{ ds^2 = \[r^2 + {1 \over 4} e^{2\tau}\ \sum_j (\beta_j
\hat x_j)^2\]
\[-d\tau^2 + {dr^2 \over r^2}\] + {r^2 e^{2\tau}\over 4}
d\Omega_{D-3}^2 + d\chi^2. }
Introducing null coordinates $v= \tau +\ln
r$ and $u=\tau - \ln r$, the metric becomes
\eqn\betterform{ -e^v[e^{-u}
+ {1 \over 4} e^u\sum_j (\beta_j \hat x_j)^2] du dv + {1\over 4} e^{2v}
d\Omega_{D-3}^2 + d\chi^2.}
Let us compare this with the asymptotic
structure of Minkowski spacetime. In null coordinates $V=t+r, U=t-r$ the
flat metric is $ds^2 = -dUdV + {1\over 4}(V-U)^2 d\Omega^2$. Near future
null infinity, $V\gg U$, so the metric is just $ds^2 = -dUdV + {1\over
4}V^2 d\Omega^2$. This is clearly very similar to \betterform. In fact we
can set $V = e^v$. In general we can also define a new $U$ coordinate so
\betterform\ is similar to Minkowski spacetime. However on the subspace
orthogonal to the rotation planes, $U =- e^{-u}$ only takes values less
than zero. Thus, in these directions, the generators of null infinity are
incomplete. This also happens for the Schwarzschild bubbles, and reflects
the fact that the exponentially expanding bubbles hit null infinity.
Now if two observers hit different points on future null infinity, then
they eventually lose causal contact with each other, although they can
both
send signals to a third observer in the interior. Since there are complete
timelike geodesics whose past does not include the entire spacetime,
these hybrid bubble
spacetimes have horizons.

One might argue that these horizons are only present for a set of
observers
of measure zero, and hence are not of much physical interest. However, we
now present a general argument that observers
near the exponentially
expanding part of the bubble are attracted to it. So it is likely that
an open set of observers have horizons. Given
any congruence of timelike geodesics with tangent vectors $\xi^\mu$, the
change in the convergence $c= -\nabla_\mu \xi^\mu$ along the geodesic
satisfies the Raychaudhuri equation \wald\ which implies:
\eqn\raych{
\dot c \ge  {c^2\over 3}  + R_{\mu\nu}  \xi^\mu \xi^\nu.}
If $R_{\mu\nu}  \xi^\mu \xi^\nu \ge 0$ (which is equivalent to a condition
on
the stress energy tensor known as the strong energy condition) then $c$
always increases along the geodesics. This just reflects the attractive
nature of gravity. In de Sitter space, the timelike geodesics orthogonal
to
surfaces of constant global time are diverging at a constant rate, i.e.,
$c$ is constant and negative. This is consistent with \raych\ since the
last term on the right is negative: a positive cosmological constant does
not satisfy the strong energy condition. However the bubble metrics we are
considering are vacuum solutions, so the last term vanishes. Since
timelike
geodesics in part of the
bubble are expanding as in de Sitter space, the only way \raych\ can be
satisfied is if the nearby geodesics in other directions are converging
toward them. Thus, nearby observers are attracted to the exponentially
expanding part of the bubble. This clearly applies to the Schwarzschild
bubble as well.

The higher dimensional Kerr metric with one angular momentum parameter
nonzero
can be dimensionally reduced along a circle to obtain a spherical brane
expanding in a fluxtube \dggh. It was previously shown that the brane
continues to accelerate outward. This is consistent with our analysis
since the brane lies on the higher
dimensional bubble, and is expanding in directions orthogonal to the
plane of rotation. Hence it does continue to accelerate. The fact
that the rest of the bubble stops accelerating
has apparently not been noticed
previously.

\subsec{Quantum stability}

The quantum stress energy that is generated in these higher-dimensional
Kerr backgrounds is expected to fall like $-1/r^{10}$ (in ten dimensions)
since the radius of the
compact direction grows asymptotically like $r$.  Thus the
equations of motion are unaltered at infinity (unlike the case of the
Schwarzschild bubble, where one has a constant energy density of order
$-1/r_0^{10}$ in the asymptotic region). As long as there are no infrared
problems (i.e. as long as we rotate at least two  planes), we expect that
one can absorb the effects of the quantum-induced stress-energy by
turning on small radial gradients of the dilaton and metric perturbations,
supported near the origin of the rotated complex planes.

This also suffices to stabilize
the orbifold \KerrIDhigherD\ without a bubble present,
for sufficiently large $R$ so that there are no tachyons.
One case of particular interest for the question
of observables is the following.  Consider the ``twisted
circle'' orbifold
\eqn\twistcirc{
(S^1\times \IC^q)/\IZ_N ~~\times ~~ \IR^{8-2q,1}.
}
Let us parameterize the complex planes by $z_1,\dots, z_q$,
the circle by $\chi$, and the Minkowski factor by $x^\mu$.
For $q>1$, the quantum stress-energy which would be generated near the
origin of $\IC^q$ in this background can be absorbed
radially as just discussed.  This produces a vacuum
which has supersymmetry broken near the origin ($z_i=0$)
of the complex planes $\IC^q$, with a small source
of stress-energy localized near $z_i=0$, but has
asymptotic local supersymmetry for large $z_i$.
By choosing appropriately
the angular momentum parameters, one can embed a
Kerr bubble into this spacetime (with rotation only in the $\IC^q$
directions).  It expands exponentially
along the $\IR^{8-2q}$ directions but stays near the origin of
the complex planes $\IC^q$ (see Fig. 5).
The eternal acceleration
of this background in the $x^\mu$ directions will
be interesting for the discussion of observables below.
\fig{Asymmetric expansion of a hybrid bubble. In the darker region,
local supersymmetry is broken to a larger extent; local
supersymmetry emerges asymptotically at large $z$. It seems that
the space-eating bubbles do not find the regions of unbroken
local supersymmetry very tasty.}
{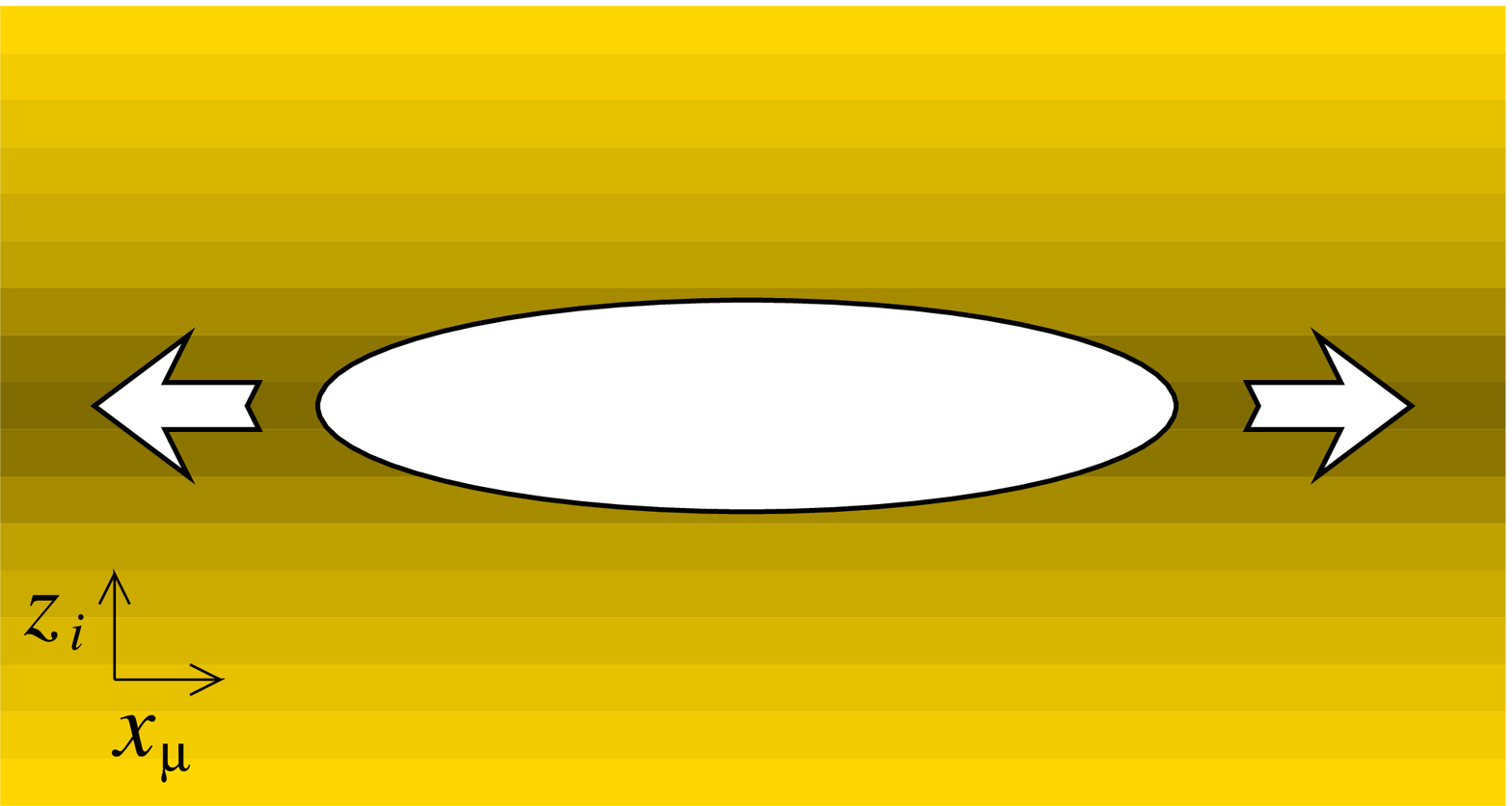}{3truein}
For Kerr bubbles with all angular momentum parameters
turned on, the bubble stops accelerating as discussed
above, and there are no horizons.  It is tempting to conclude
from this that the stabilization of the bubble geometry
(which motivated our study of the Kerr bubbles as opposed
to the Schwarzschild bubbles) removes the horizons, corroborating
the point of view in \smatrix.

However, the hybrid cases discussed above with eternal acceleration in a
subset of the directions, provide one set of apparently perturbatively
stable backgrounds without a standard S-matrix.  Observers on the bubble
separate exponentially in the $x^\mu$ directions, and lose causal contact
with each other.  They can send information off the bubble to a third
observer, but they cannot send information to each other and no single
observer can accumulate all the data in the full S-matrix.

We now consider possible nonperturbative quantum instabilities.
Nonperturbatively, the backgrounds with all angular momentum
parameters nonzero, are probably also stable against nucleation of
additional bubbles. Since the compact direction opens up asymptotically,
far from the first bubble the spacetime does not have the right
boundary conditions to nucleate another bubble. This is roughly because
the spacetime is essentially flat space asymptotically, which is stable.
However, in the hybrid cases with horizons, there are directions where
the compact direction remains finite, and the spacetime resembles the
Schwarzschild bubble. In this case,  there appears to  be no reason why
additional bubbles could not appear far from the first in these particular
directions. If so, it would be interesting to investigate this further,
studying what happens when the bubbles collide, and checking whether any
horizons remain.

\newsec{Particle Creation in Quantum Field Theory}

A basic phenomenon which arises in generic time-dependent
backgrounds is particle creation. This is particularly simple to
study in backgrounds like the bubbles we described above with all
angular momentum parameters nonzero,
which have a Minkowski-like null infinity region
where we can unambiguously define in and out vacua for massless
fields.  Because there is no global time translation symmetry,
an initial positive frequency mode generally
turns into a linear combination of positive and negative frequency
modes in the future (and vice versa).  In the Kerr bubble
solutions, we have phases of mild time dependence
asymptotically in the past and future (the ``Milne'' epochs),
interrupted by the
``de Sitter" epoch of exponential contraction and expansion of the bubble.
Intuitively, we expect that at least for a long enough de Sitter
phase, we should find the particle creation dominated by this
epoch; this is confirmed in our analysis below.

Here we will study this process in detail for a massless scalar
field
\foot{ Note that the choice of coupling of $\phi$ to the
metric will not be important because the scalar curvature
vanishes in our backgrounds.}
$\phi$
in the four-dimensional Kerr bubble geometry \KerrbubbleIV; we expect it
to work similarly in higher dimensions with generic rotation
angles, where this analysis would apply, for instance, to the dilaton
in string theory. Similar particle-creation will also arise for other
fields (like the graviton) but we will not discuss it here.
Our goal is to express the future Minkowski modes in terms
of the past Minkowski modes and to calculate the Bogolubov
coefficients relating them.

One way to do this is to explicitly solve for the wave-functions of
the field, by solving equations like \eom\ and its generalization to
other backgrounds, and comparing the positive-energy modes at past
null infinity with the positive-energy modes at future null infinity.
However, solving these equations in general seems to be too complicated.
Thus, we will follow another strategy,
following \hawking, which is to consider
frequencies $\omega > \! \! \!> 1/r_0$ and make a geometric optics
approximation.
In other words, for each mode we will use the fact that
the phase is (approximately) constant along each geodesic.
\fig{Pair of geodesics in the four dimensional Kerr bubble geometry used
for the particle creation calculation.}{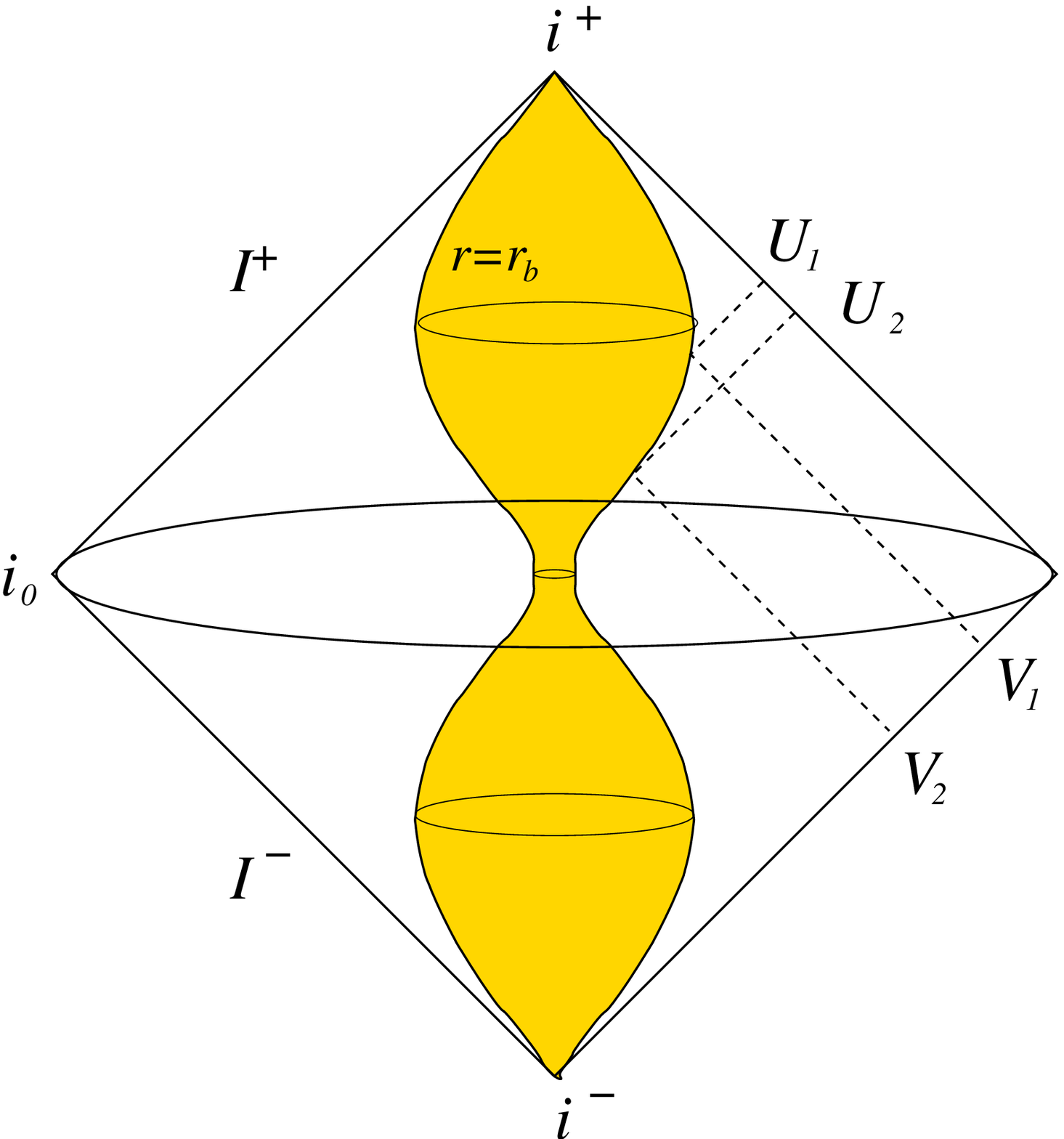}{4truein}
Near future null infinity, the phase of a future Minkowski
mode is simply proportional to the coordinate distance between two
geodesics. Following these geodesics to the past in the full geometry,
we will find the phase as a function of the Minkowski coordinates near
past null infinity, giving us the Bogolubov transformation.

More specifically, consider an s-wave mode, or some other low
angular momentum mode. Decomposing this mode into a spherical
harmonic and a function independent of the angular variables, one
can reduce the problem to two dimensions. If the  angular momentum
is small, it will not affect  the effective 2d wave equation much
and the mode will behave similarly to an s-wave. From now on we
will express all quantities in the two-dimensional language.

We will be interested in an s-wave mode at ${\cal I}^+$ of purely
positive frequency,
\eqn\futuremode{ \phi \sim e^{i\omega U},}
with respect to a lightcone coordinate $U=T-X$, where $X$ and $T$
are the usual Minkowski coordinates in the radial and time
directions at infinity. For this mode, the phase
difference between two geodesics located at $U_1$ and $U_2$ will be
$e^{i \omega \Delta U}$, where $\Delta U\equiv U_1-U_2$ is the
coordinate distance between the two geodesics (labelled 1 and 2 in
Fig. 6) on ${\cal I}^+$. By solving the geodesic equation and tracing
geodesics 1 and 2 back to ${\cal I}^-$, we can trace the phase difference
between the values of $\phi$ on the two geodesics back to
\eqn\pastmode{ e^{i\omega\Delta U(\Delta V)}, }
where again $\Delta V\equiv V_1-V_2$ is the coordinate distance
between the two geodesics on ${\cal I}^-$, and $V \equiv X+T$.
This form reflects the
fact that the mode has a constant phase (in the geometric optics
approximation), but its expression in terms of $U$ can be traded
for that in terms of $V$ on ${\cal I}^-$ by determining the
distance between our two geodesics on ${\cal I}^-$.

For the four dimensional Kerr bubble \KerrbubbleIV, we can determine the null
geodesics relevant for the s-waves by setting
\eqn\geodcond{
ds^2\equiv 0 \to
d\tau^2={{dr^2}\over{r^2-\beta^2-r_0r}}. }
Integrating
this condition, we find the relation between $r$ and $\tau$
satisfied along the null geodesics:
\eqn\geodesics{ e^\tau =
e^{-C}\biggl(r-{r_0\over 2}\pm\sqrt{r^2-\beta^2-r_0r}\biggr), }
where the integration constant $C$ labels the geodesic, and where
the $\pm$ comes from the two possible signs when taking the square
root of \geodcond, and distinguishes outgoing and incoming geodesics.

Since we are interested in the modes \futuremode\ at
null infinity, we need to know the null coordinates $U,V$ in terms
of $r,\tau$.  They are given by
\eqn\nullcoordU{ U=-re^{-\tau}+{\beta^2\over 4}{e^\tau\over r}, }
\eqn\nullcoordV{ V=re^{\tau}-{\beta^2\over 4}{e^{-\tau}\over r}. }
In terms of $U,V$, the metric for large $r$ ($r\gg r_b,\ r\gg \beta,\
r\gg r_0$) and large $|\tau|$ reduces to (for $d\chi=d\phi=0$)
\eqn\minknull{ ds^2=-dUdV. }
Plugging \geodesics\ (with the plus sign) into \nullcoordU\ we
find the future asymptotic $U$ coordinate of the null geodesics as
a function of $C$:
\eqn\Ugeod{ U\to {1\over 2}(-e^C+\beta^2 e^{-C}). }
Similarly, the past asymptotic $V$ coordinate of the geodesics is
\eqn\Vgeod{V\to e^{-C}K_0-{\beta^2\over 4}{e^C\over K_0},}
where $K_0\equiv {\beta^2\over 2}+{r_0^2\over 8}$.

Let us fix the geodesic 1 in the figure to be a reference geodesic
with a given value of $C$, $C=C_1$.  We would like to solve for
$U_2$ in terms of $V_2$.  To start, let us solve for $e^{C_2}$ in
terms of $V_2$ using \Vgeod, and then plug the result into \Ugeod.
We find
\eqn\gammV{e^{C_2}={{2K_0}\over\beta^2}\biggl(-V_2+
\sqrt{V_2^2+\beta^2}\biggr).}
This leads to
\eqn\Usoln{\eqalign{ \Delta U= U_1-U_2= & {1\over
2}\biggl({{2K_0}\over\beta^2}\bigl(-V_2+\sqrt{V_2^2+\beta^2}\bigr)
-e^{C_1}\biggr)\cr & +{\beta^2\over 2}\biggl(e^{-C_1}-
{1\over{{2K_0}\over\beta^2}\bigl(-V_2+\sqrt{V_2^2+\beta^2}\bigr)}\biggr).\cr}}
It is worth remarking that in the limit of the Schwarzschild bubble,
$\beta\to 0$, this reduces to
\eqn\UsolnSchwarz{\Delta U^{(\beta=0)}={1\over
2}\biggl({r_0^2\over{8 V_2}}-e^{C_1}\biggr).}

Plugging \Usoln\ into \pastmode, we would like to project the result onto
Fourier components $e^{\pm i \omega' V} \sim e^{\pm i\omega' V_2}$
(recall that we are keeping $V_1$ fixed), in order to read off the
Bogolubov coefficients which determine the extent of particle
creation in our background.  From the nontrivial form of \Usoln,
we already see that our pure positive frequency mode in the future
\futuremode\ does not continue back to a pure positive frequency
mode in the past, and thus there is indeed particle creation.

More explicitly, from the standard definition of Bogolubov
coefficients it follows that
\eqn\bogtranf{ {1 \over \sqrt{\omega}} e^{i\omega\Delta U}=\int
d\omega^\prime \biggl(\alpha^*_{\omega^\prime\omega} {1 \over
\sqrt{\omega'}} e^{-i\omega^\prime V_2}
-\beta_{\omega^\prime\omega} {1 \over \sqrt{\omega'}}
e^{i\omega^\prime V_2}\biggr). }
So, we can extract the Bogolubov coefficients $\alpha_{\omega'
\omega },\beta_{\omega' \omega }$ as
\eqn\Bogalpha{\alpha_{\omega^\prime\omega}={1 \over 2\pi
}\sqrt{\omega^\prime\over\omega}\int_{-\infty}^\infty dV
e^{i\omega^\prime V}e^{i\omega\Delta U(V)}, }
\eqn\Bogbeta{\beta_{\omega^\prime\omega}={1 \over 2\pi }
\sqrt{\omega^\prime\over\omega}\int_{-\infty}^\infty dV
e^{-i\omega^\prime V}e^{i\omega\Delta U(V)}, }
where $\Delta U(V)$ is given explicitly by \Usoln, and where
we are ignoring overall phases since the quantities we are interested in,
such as the number of particles produced, are determined by the
magnitude of $\beta_{\omega^\prime\omega}$.
For a fixed mode $\omega$,
the total number of particles produced (in a range between
$\omega$ and $\omega+d\omega$) is given by
\eqn\partN{dN_{\omega}=d\omega \int_{1/r_0}^{M_P}d\omega^\prime
|\beta_{\omega^\prime\omega}|^2,}
where we have restricted the integral to our regime of validity,
the lower end coming from the validity of the geometric optics
approximation and the upper end from the inapplicability of
quantum field theory techniques at the Planck scale.

 From these exact results we would like to extract certain
qualitative features of the physics.  First, we would like to
check that the particle creation is not large enough to back-react
significantly on the geometry.  Second, we would like to
understand where (or when) the bulk of the effect arises, and to
test our intuition that the particle creation is dominated by the
de Sitter epoch in the Kerr bubble spacetime.

In order to accomplish this, let us consider first a stationary
phase approximation to the integrals \Bogbeta,\Bogalpha. In the
Schwarzschild bubble, we have $\beta = 0$ and thus $V\gg \beta$
for any nonzero $V$. Let us check whether the stationary phase
point of the full solutions \Bogbeta,\Bogalpha\ is in this regime.
If so, then the de Sitter phase indeed dominates the particle
creation. In the limit where $V\gg \beta$  and
${\beta^2\over r_0^2}\ll {\omega\over\omega'}$ (the Schwarzschild
bubble approximation, whose self-consistency condition we will
discuss momentarily), the stationary phase is at
\eqn\statphase{V^2 = V_s^2\equiv \pm {K_0\over
2}{\omega\over\omega^\prime},}
for $\alpha_{\omega\prime \omega}$ and $\beta_{\omega\prime
\omega}$ respectively (namely, the phase $V_s$ is real for
\Bogalpha\ and imaginary for \Bogbeta; in the latter case we must
deform the contour to go through the stationary phase point).
In order for this approximation to be
self-consistent, from \statphase\ and the definition of $K_0$
above we need
\eqn\selfcons{{\omega\over\omega^\prime}\gg {\beta^2\over r_0^2}.}
The smallest ratio $\omega/\omega^\prime$ that we can discuss
reliably is $1/(r_0M_P)$.  We can thus choose $\beta$ small enough
so
that \statphase,\selfcons\ are satisfied for all the frequencies we consider.
This regime of very small $\beta$ is of interest
in any case for producing a long de Sitter phase.

Plugging the stationary phase \statphase\ into the integrands
\Bogalpha,\Bogbeta, we arrive at the stationary phase approximation
to the Bogolubov coefficients:
\eqn\statBogalpha{\alpha_{\omega^\prime\omega}\sim
e^{i{r_0\over
2}\sqrt{\omega^\prime\omega}},}
\eqn\statBogbeta{\beta_{\omega^\prime\omega}\sim
e^{-{r_0\over
2}\sqrt{\omega^\prime\omega}}.}

In the Schwarzschild bubble limit (which, as just discussed, is a
good approximation to the Kerr case as well for
small enough $\beta$) one can evaluate the integrals \Bogalpha,\Bogbeta\
exactly in terms of Bessel functions.  The results are
\eqn\AlphaTwo{|\alpha_{\omega' \omega}| = {r_0 \over 8}
H_{-1}^{(1)} \left({1 \over 2}r_0 \sqrt{\omega \omega'}\right)}
(where $H$ is a Hankel function) and
\eqn\BetaTwo{|\beta_{\omega' \omega}| = |\alpha_{\omega'
(-\omega)}| = {r_0 \over 4\pi} K_{1} \left({1 \over 2}r_0
\sqrt{\omega \omega'}\right)}
(where $K$ is a modified Bessel function); these have the same
behavior as \statBogbeta\ and \statBogalpha\ in an asymptotic
expansion.

 From \statBogbeta,\BetaTwo\ we see that the nontrivial Bogolubov
coefficient $\beta_{\omega' \omega}$ dies to zero exponentially for large
frequency, so that one expects
\foot{Note however that the relation between particles and
the energy-momentum tensor is usually very complicated
in a time-dependent spacetime.}
the produced particles to carry finite energy, in contrast to the
situation in the models of \vijay\ and the non-Euclidean conformal
vacua in \dSCFT, and similarly to the finite-energy spacelike
branes in \spacebranes. The particle creation in our backgrounds
is soft at high frequency, and is therefore consistent with our
QFT analysis. Because of this exponential suppression, the total
number of particles produced, determined from \partN\ by
integrating over $\omega$ (with the IR cutoff $1/r_0$), is finite.
Since these particles can spread over the infinite region of null
infinity, the energy density produced is vanishingly small and the
particles we produced do not back-react significantly on the
geometry.

It is interesting to note that the Bogolubov coefficients
\AlphaTwo\ and \BetaTwo \ are of exactly the same form as those
obtained for a mirror moving with a constant proper acceleration
(i.e. along a hyperbolic trajectory) in flat space \BD. So there
is a close analogy to the previously studied moving mirror problem.
However,
whereas the moving mirror was an ad hoc construct, requiring
an external force to keep the mirror accelerating, the bubble spacetimes
are solutions to Einstein's equations.

In summary, within the range of frequencies between $1/r_0$ and $M_P$, we
find
a rich spectrum of particles produced by the background, but not
enough to significantly back-react on the geometry.  In the next
section we will discuss the string theoretic description of
particle creation in general terms, which will apply in particular
to our bubble geometries.

\newsec{Particle Creation in String Theory and Non-local String Theories}

Most of our analysis, including the computation of particle
creation we just completed in \S4, has been in the framework of
low-energy effective quantum field theory and general relativity.
In this section we will move on to string theory. In \S5.1, we
will present an observation concerning the description of particle
creation in string theory, in particular showing that Nonlocal
String Theory (NLST) \NLST\ arises naturally on the worldsheet in
describing the resulting squeezed states. This assumes that there
is a generalization of string perturbation theory to time
dependent backgrounds -- perhaps in backgrounds such as our
bubbles this generalization proceeds via translation from the
Euclidean continuation. Such a generalization has yet to be
formulated however, and we discuss some of the challenges involved
in doing so in \S5.2.

\subsec{Particle creation in string theory}

To describe gravitational particle creation from the string theory
point of view, we will use general methods known from quantum
field theory in curved space and translate them into perturbative
string theory.

One way to formulate particle creation is to look (in the
Heisenberg picture) at a vacuum state
$|in\rangle$ which does not have any particles in it;
i.e. it is killed by the annihilation operators
$a_{in}$, obtained from a mode expansion of the field
which reduces in the far past to an ordinary Fock space expansion in which
creation operators $a^\dagger_{in}$ multiply pure positive
frequency modes and annihilation
operators $a_{in}$ multiply pure negative frequency
modes. In general, the modes multiplying $a^\dagger_{in}$
and $a_{in}$ do not reduce in the far future to pure positive
frequency and pure negative frequency modes, respectively. That is,
the state $|in\rangle$ is not the same as the state
$|out\rangle$ killed by the operators $a_{out}$ multiplying
the pure negative frequency modes of the fields in the future.
Instead, it is a
squeezed state $|\Psi_\kappa\rangle$, which can be written in terms of
a basis of out-going oscillators ($a^I_{out} | out\rangle = 0$)
as
\eqn\basicev{|\Psi_\kappa\rangle \equiv
|in\rangle = C e^{\kappa_{IJ} a^{\dagger I}_{out}
a^{\dagger J}_{out}}|out\rangle,}
where $C$ is a normalization constant (these states are
normalizable for sufficiently small $\kappa$), and
$\kappa=-{1\over 2}\beta\alpha^{-1}$ in terms of the matrices
$\alpha,\beta$ of Bogolubov coefficients defined as in \bogtranf.

When we calculate S-matrix elements using the bra state
$\langle \Psi_\kappa | =\langle in |$ (and oscillator
excitations above it) in the future, this amounts to treating
the particles produced as part of the background.


Another way to phrase particle creation is by looking at
correlation functions in the ``empty" $|in\rangle$ and
$|out\rangle$ vacua. The above discussion makes it clear that in
general time-dependent backgrounds correlation functions of the
form
\eqn\creation{\langle out|a_{out,1} \cdots a_{out,2n} \ |in\rangle}
or
\eqn\creationtwo{\langle out|a^{\dagger}_{in,1} \cdots
a^{\dagger}_{in,2n}\ |in\rangle}
will be non-vanishing, and will correspond to the probability for
finding $2n$ outgoing particles when starting from the initial
vacuum state (or vice versa). On the other hand, the correlation
function similar to \creationtwo\ with $\langle in | = \langle \Psi_\kappa
|$
appearing instead of
$\langle out |$ will vanish for $n > 0$.
In our discussion we will be assuming that one can
construct asymptotic scattering states for the background of
interest.  We are also assuming that interactions can be neglected,
so that the free field result for the particle creation is
a good approximation, as in \S4\ where the coupling
can be made arbitrarily small and the particles created
have a vanishing number {\it density}.
In general, the fields in the Heisenberg
representation, or the states in the Schr\" odinger representation,
will evolve nontrivially and leave us with something more
complicated than the simple pure squeezed state in \basicev.
In situations with an ambient temperature, the interactions
would be expected to lead to thermalization, and in general
one expects decoherence to occur effectively for observers not privy
to the global structure of the state.

In string theory, we expect that as in flat space, the creation
and annihilation operators for the asymptotic scattering states
will be related to vertex operators $V_{\pm}^{in,out}$, where the
sign depends on the sign of the energy with respect to the
corresponding vacuum. We expect in cases which have an S-matrix
(such as our Kerr bubbles with all rotation parameters non-zero),
that just as in flat space, S-matrix elements between ket states
created by the $a_{in}^\dagger$'s on the $| in \rangle $ vacuum
and bra states created by $a_{out}$'s on the $\langle out|$ vacuum
correspond to worldsheet correlation functions of the
corresponding integrated $V_{+}^{in}(z)$ and $V_-^{out}(z)$ vertex
operators. Just as in field theory, the in and out bases are not
independent, and we can express the in vertex operators as linear
combinations (Bogolubov transformations) of out operators and vice
versa.

The correlation functions we compute in field theory depend on the
initial and final states we use, so the results we expect to find
in string theory depend on which initial and final states we have
there. If we start from the Euclidean string theory (in cases for
which such Euclidean versions exist), the analytic continuation
defines a particular state which we can call the
``Euclidean vacuum" $| Euclidean\rangle $.
Generally, this is not the same as the
natural Lorentzian vacua with no particles (see, e.g.,
\refs{\sussfischler,\dSCFT}), but rather it looks like a squeezed
state. It seems that correlation functions in Euclidean string
theory will naturally continue to Lorentzian correlation functions
in this particular state. It is interesting to ask how we can
compute in string theory correlation functions in the usual empty
Lorentzian vacua $|in\rangle$ and $|out\rangle$,
or in particular interesting squeezed states
like $\langle \Psi_\kappa|$.
We will argue that to do this we need to
deform the worldsheet action by non-local terms, as in \NLST.

In order to see this, let us first recall how to describe
different initial and final states in a path-integral formulation
of field theory. If (staying in the Heisenberg
representation so that the fields evolve but the
states do not) we consider states which are eigenstates
of the field $\phi(T_i)$ in the asymptotic past and
$\phi(T_f)$ in the asymptotic future
(for string theory we will be interested in
the limit $T_{f,i}\to \pm \infty$), we have the relation
\eqn\basicpath{ \langle\phi(T_f)=\phi_b |\phi(T_i)=\phi_a\rangle
= \int
[d\phi]\bigg|_{\{\phi(t=T_i)=\phi_a, \phi(t=T_f)=\phi_b\}}e^{iS}, }
so the boundary conditions in the path integral language
correspond to the choice of states appearing in the matrix element in
operator language. If we want to consider more general
wavefunctions $|\Psi_{a,b}\rangle $ instead of
$|\phi_{a,b}\rangle$, then we simply decompose them in the
$|\phi\rangle $ basis :
\eqn\decomp{|\Psi\rangle = \int [d\phi]|\phi\rangle \langle\phi
|\Psi\rangle, }
giving
\eqn\decompPI{\langle \Psi_b | \Psi_a\rangle =
\int[d\phi_a]\int[d\phi_b]\int[d\phi]\bigg|_{\{\phi(T_i)=\phi_a,
\phi(T_f)=\phi_b\}}
\Psi_b^\dagger [\phi(T_f)]\Psi_a[\phi(T_i)]e^{iS}.}

We are interested in particular in wavefunctions $|\Psi\rangle$
which are squeezed states.  Above we wrote such squeezed states in
terms of creation operators \basicev, but we can equivalently
write them in terms of the fields as Gaussian wavefunctions of the
form
\eqn\state{\Psi_s \sim C e^{\tilde c\int d^{d-1}\sigma
d^{d-1}\sigma'
\phi_+(\sigma)\Delta(\sigma,\sigma')\phi_+(\sigma')}, }
where $\tilde c$ is a constant, $\sigma$ is a coordinate on the
boundary $\del$ in terms of which we define the squeezed state
(this boundary can be space-like or null), and $\phi_{\pm}$ are
the positive and negative frequency parts of the field at that
boundary. We have taken into account the fact that in general the
wavefunction will not be local on $\del$ (see, e.g., \dSCFT).

To describe matrix elements between these states, we can plug them
into \decompPI.  Since the wavefunctions \state\ inserted in the
path integral in this case are exponentials of the form
$\Psi_s=e^{W[\phi]}$, we can reinterpret the log of these
insertions as relatively simple contributions to a boundary
action. In general these actions will not be local on the
boundary, as expected since squeezed states embody long-range
correlations, but in some special cases the boundary action may be
local.

Having obtained the effect of the squeezed states as a shift in
the boundary action, we can now deduce the string theory
description. We can consider the boundary action we just derived
as part of the interaction Hamiltonian, and treat it
perturbatively in the parameter $\tilde c$ appearing there.
Bringing down powers of the boundary action into correlation
functions leads to contributions to amplitudes in which those
boundary fields are contracted with fields in other vertices in
the diagram. This has the same effect as adding external
(integrated) vertex operators in string theory. Thus, the squeezed
state of a space-time field $\phi$ is reproduced in string theory
by introducing a shift in the worldsheet action by the
corresponding multilocal function of integrated vertex operators
corresponding to $\phi$. Note that we avoided issues of operator
ordering by writing the squeezed state purely in terms of positive
frequency modes.

For instance, in the case of \basicev, a correlation function
involving $\langle \Psi_{\kappa}|$ would be described by
\eqn\wsac{\delta S_{ws}=\sum_{I,J} (-{1\over
2}\beta\alpha^{-1})_{IJ} \int d^2z_1 V_+^{out,I}(z_1) \int d^2z_2
V_+^{out,J}(z_2)+\delta S_{E, in}+\delta S_{E, out}}
which is manifestly non-local on the worldsheet\foot{The action is
not real, but this should not be surprising when we have particle
production and we are looking at a complex final state.}.
Here we have included terms $\delta S_{E, in}+\delta S_{E, out}$
describing the (nonlocal) shift from the Euclidean vacuum to the
empty Lorentzian vacua, which are
applicable if we formulate the string theory using the Euclidean
continuation.  The
appearance of
worldsheet non-locality here (and its connection to a boundary
action) is very similar to its appearance in
the discussion of multi-trace operators, related to multi-particle
states in AdS, in \NLST\ (and its interpretation in
supergravity on AdS in terms of a boundary action \NLSTboundary).
In general the sum in \wsac\ will be
replaced by an integral over all possible outgoing modes (if we
are describing the final state, or over incoming modes if we are
describing the initial state). The form of the deformation above
is valid when its coefficient is very small, in which case the
deformation manifestly preserves worldsheet conformal invariance;
for a finite coefficient the form of the deformation would generally
receive corrections, corresponding to the backreaction of the
state on the background. Our discussion here was for the case of a
single free scalar field (such as the dilaton), but there is no
problem (in principle) in generalizing it to other cases.

Using this formalism, we can translate a given initial or final
squeezed state into a non-local contribution to the worldsheet
action. Given a string theory for the backgrounds we discussed
above, we can explicitly write down the deformation we would need
for describing the natural final squeezed state
there (assuming that our original string theory described the
``empty'' vacuum state).
For example, at leading order in $\alpha'$, the string
worldsheet action in our bubble backgrounds shifts to include a
term of the form \wsac, with $\alpha_{\omega\omega^\prime}$ and
$\beta_{\omega\omega^\prime}$ as given in
\S4, and with the massless vertex operators at leading order given by
the solutions of the wave equation (such as \eom) for the
corresponding spacetime fields. We expect that at low energies the
string theory results derived from these actions
will reproduce field theory correlation
functions in the appropriate initial and final states.

As another example, if we have a string theory describing a
background including $dS_3$, and
we want to do computations in the final squeezed state (which is
the time-evolved initial vacuum state for a particular scalar
field on $dS_3$), then using the results of \dSCFT,
we need a similar nonlocal deformation of the worldsheet action.
Interestingly, the dS/CFT
conjecture would relate these deformations to double-trace
deformations in a dual CFT, which is similar to the way the NLST
deformations were discovered in the context of double-trace
deformations of AdS/CFT \NLST. One should, however, keep in mind
that for these squeezed states there is a large backreaction on
the geometry. Also, it appears hard to define the
gravity counterpart of the exact CFT correlators, taking into account
non-perturbative aspects of gravity in de Sitter space
\dysonetal.

In the AdS/CFT context, it was shown in \NLSTboundary\
that one could simply describe the NLST deformation
in a formal supergravity approximation as
a deformation of the boundary action for the fields
in AdS \foot{It is worth
emphasizing that this boundary action is local in the AdS
coordinates but it is nonlocal on the
compact part of space, and that the worldsheet description is
nonlocal.}.
This boundary action, on the timelike boundary of AdS, implements
deformed boundary conditions for the supergravity fields.
In our case here, we have a boundary action corresponding to
the squeezed state we consider in the future, on null or spacelike
infinity.  Both types of boundaries
lead to an NLST deformation on the worldsheet.

Thus, NLSTs arise naturally whenever we want to compute
correlation functions in states which are not the natural initial
and final states of string theory (e.g. the states defined by the
Euclidean continuation). This does not mean that we {\it have} to
use NLST for time dependent backgrounds; if we calculate matrix elements
(or
equivalently the corresponding path integral) involving the
``Euclidean vacuum'' state
$\langle Euclidean |$ instead of the squeezed state
$\langle \Psi_s |$ we would use ordinary ``local" string theory
rather than NLST. However, in many contexts it may be natural to
use a final state squeezed with respect to
$\langle Euclidean |$ (at least before taking into account the
effects of decoherence), and then NLST seems to be required.
Generally NLST seems to be required even to describe the natural
Lorentzian vacua, if we use a continuation from Euclidean space
to formulate string perturbation theory, since in time-dependent
backgrounds the Lorentzian vacua are related by a non-trivial
Bogolubov transformation to the ``Euclidean vacuum". The S-matrix
is presumably related to correlation functions of the vertex
operators $V_{\pm}^{in,out}$ in the Lorentzian vacua $\langle out|$
and $|in\rangle$. If we can
formulate string theory directly in Lorentzian space, then the
local string theory may correspond to doing computations in the
natural Lorentzian in and out vacua (though this is not obvious),
but we would need NLST to do computations directly in squeezed
states, such as the ones arising by particle creation from the
vacuum. Note that the formulation we described in this section
applies also to Minkowski space, so we could choose also there
non-trivial (correlated) initial and/or final states, which would
be described by a non-local theory on the worldsheet. We see that
worldsheet non-locality can arise in much more general contexts
than anti-de Sitter space where it was first discussed \NLST; it
would be interesting to understand more generally when such
non-locality occurs, and how to quantize and do computations in
these non-local theories (beyond perturbation theory in the
non-local deformation).

In our discussion here we generally ignored the
backreaction of the initial or final state on our background.
Obviously, when squeezed states with many particles are involved,
this backreaction may not be negligible and needs to be taken into
account. This may change our conclusion that any choice of state
is allowed, and is realized by a particular non-local deformation
of the worldsheet action.  As we saw in \S4, in our bubble geometries
the back-reaction is small.

A somewhat confusing aspect of the discussion here is that even if
we compute with the local worldsheet theory, it seems that 2-point
functions of incoming or outgoing vertex operators on the sphere
should be non-zero. Usually in string theory, 2-point functions
vanish on the sphere because of the infinite volume of the group
of conformal transformations preserving two points, except when
there is another infinity to cancel this (as in for example
\SLC). In our case it is not clear where such an additional
infinity would come from, though conceivably it could come
from integrating over the time direction.

\subsec{Time-dependent backgrounds in string theory}

In general, the description of time-dependent backgrounds
(without any light-like Killing vector) in string theory is
a notoriously difficult subject.
(String theory on plane wave backgrounds, which are time-dependent but
independent of a light-cone time direction, was studied in many papers
starting with \dalsi.)
One longstanding problem \HorowitzAP,
which is starting
to receive more attention
\singrefs, is that most known
time-dependent backgrounds, and in particular the ones which are
relevant for cosmology, have singularities. Such singularities
pose an obvious problem for low-energy gravity computations, but
it may be that these are resolved in string theory either classically or
quantum mechanically and pose less of a problem there.

A complementary direction is to avoid the singularities by considering
for example de Sitter space or the bubble solutions we have
studied in this paper, which allow one to focus on the many
other important questions in string cosmology.
Even the study of non-singular time-dependent
backgrounds is
problematic in string theory. Among the unsolved problems are
how to deal with cosmological horizons, how to prove a no-ghost theorem,
and how to define physically meaningful observables.
An additional problem is that string perturbation theory
as we know it currently is only
well-defined either in light cone gauge (which is
not available in generic time-dependent backgrounds)
or in Euclidean space -- both on the worldsheet
and in spacetime -- where one has the usual genus
expansion.

The simplest way to try to define Lorentzian string theory may be
by analytic continuation of results from Euclidean space.
For example, the bubble geometries we study here, which
are double analytic continuations of black hole solutions,
have Euclidean continuations
which have been well studied in the black hole context \gpy.  However,
some time-dependent backgrounds do not have a straightforward
Euclidean continuation, so it is not clear how to even formulate
string perturbation theory in such backgrounds.
Furthermore, even in cases such as ours which do have a Euclidean
continuation, there are issues involved in translating Euclidean computations
to Lorentzian ones, as we will see shortly.

One of the first issues we need to address when formulating string
theory on a bubble background of the type discussed above, is
the question of
whether the leading order solution in the $\alpha^\prime$ expansion
extends to a full solution of classical string theory.
Unlike some other time-dependent backgrounds like Lorentzian orbifolds and
the Nappi-Witten background \nappiwitten, our backgrounds are not exact
CFTs. However, they are solutions to Einstein's equations so they
are conformal at 1-loop order, and there are no singularities
or regions of strong coupling so both the $\alpha'$ expansion
and the string loop expansion are good everywhere in the
spacetime.  As we discussed in \S3, the instabilities introduced
by the string loop expansion (tadpoles introduced by quantum
mechanically generated stress energy) are localized in space
in more than two directions, and thus can be absorbed by mild
radial variations of the supergravity fields.  The corrections
to the field equations arising from $\alpha^\prime$ effects will
also be localized, and may be absorbable similarly in small radial
variations of the supergravity fields.  In particular,
in analyzing massless scalar fields (and gravitons in
the four-dimensional Schwarzschild case)
in our study of classical stability in general relativity,
we found no massless (or tachyonic) modes localized near
the bubble.  If this persists to hold for all massless fields in
all examples, then the solutions are isolated, and
$\alpha^\prime$-induced tadpoles will shift slightly
but not destabilize the solution.

On the worldsheet, this question of classical
stability amounts to the following.
We solved the $\beta$-function equations to 1-loop order in
$\alpha^\prime$ by solving Einstein's equations.
At higher loop orders we expect to
find non-zero beta functions which will induce some flow, and the
question is whether this flow has a fixed point which is close to
the original background (with corrections of the order of the
curvature in string units).
It is natural to study this question in Euclidean space, where
we expect the worldsheet CFT to be unitary. The Euclidean
continuation of our backgrounds is the same as the Euclidean
continuation of the Schwarzschild and Kerr black hole backgrounds,
so the existence of classical (Euclidean) string theory in our
backgrounds is equivalent to the existence of classical string
theory on the black hole backgrounds. As in the bubble backgrounds,
in the Lorentzian
black hole backgrounds we expect an isolated solution and no
instabilities from $\alpha^\prime$ that cannot be absorbed
in small radially-dependent shifts of the fields \CallanHS.

There are, however, some difficulties with implementing
this program.
One immediate issue is that not all the
vertex operators we would like to have in the Lorentzian case
exist in the Euclidean case, since in the latter solution there
are angular coordinates with a discrete spectrum that become
noncompact (time) dimensions in the Lorentzian continuations.

Another issue is the fact that the Euclidean continuation
has a negative mode \gpy\ related to the negative specific
heat of the corresponding black hole.  This has two interesting
consequences.

First, this mode translates into a relevant operator
on the Euclidean string worldsheet CFT, so that the
Euclidean black hole target space is an unstable fixed
point with respect to variations of the corresponding coupling
in the worldsheet sigma model.  A generic RG flow would lead
us far away from this fixed point.
However, since the solution
is isolated (there being no marginal deformations),
the $\alpha'$ corrections will only shift the unstable fixed
point slightly, and we can fine tune the worldsheet
theory order by order in $\alpha'$ to remain at the unstable
fixed point.  This fixed point will describe
a background only slightly shifted
from the leading order in $\alpha'$ general relativistic
solution we have been working with in the bulk of this paper.

Second, this negative mode leads to a divergent
one-loop path integral (genus one amplitude)
in the Euclidean continuations, even though
the corresponding black holes and bubble solutions do not have tachyonic
instabilities (at least in $D=4$).  It may be possible to interpret
this divergence as follows\foot{We have enjoyed useful discussions
with T. Banks, M. Berkooz, D. Kutasov, and A. Lawrence on this issue.}.
One may be able to analytically continue
this divergent computation to obtain a finite answer, at the
cost of introducing an imaginary part to the amplitude.
In a theory with the ordinary unitarity relations, the
imaginary part of the one-loop amplitude would be equal
to the square of the amplitude to produce pairs of particles.
In time-dependent backgrounds, we do not know the appropriate
generalization of the cutting rules, but since one can obtain
unitary evolution formally for quantum fields on
this space (with a time dependent Hamiltonian)
one may be able to make this argument precise including the effects
of the time-dependence.  If so, then the analytically continued
one-loop vacuum amplitude could provide an alternate means to
calculate the particle creation amplitude in the full
string theory, generalizing the results of \S4.

As in the case of static backgrounds, developing explicit
controlled solutions
such as those we have studied here is an important
prerequisite to addressing
these very basic questions about formulating string
theory in time-dependent backgrounds.
With such a formulation in hand, we could answer interesting
questions about how UV-sensitive quantities behave in
time-dependent string backgrounds (for example the question of
whether string theory softens particle creation
effects at high energies), and perhaps we would get a better
handle on classical singularities and on more realistic backgrounds.

\vskip 1cm
\centerline{\bf Acknowledgements}
We would like to thank T. Banks for useful and interesting discussions
during his collaboration at an early stage of this
work.  We would also like to thank M. Berkooz for many discussions
on particle creation and the relation to NLST.  In addition, we
enjoyed discussions on these and related subjects with
A. Adams, R. Emparan, S. Giddings, P. Ho\v rava,
S. Kachru, N. Kaloper, A. Lawrence, A. Linde, J. Maldacena, L. Motl,
J. Polchinski, S. Shenker, and L. Susskind.
O.A. is the incumbent of the Joseph and Celia Reskin career
development chair, and he would like to thank SLAC and the Isaac
Newton Institute for hospitality during the course of this project.
The work of O.A. and E.S. is supported in part by the
Israel-U.S. Binational Science Foundation. The work of O.A. is supported
in part by the IRF Centers of Excellence Program, by the European RTN
network HPRN-CT-2000-00122, and by Minerva.
The work of G.H. was supported in part by NSF Grant PHY-0070895.
The work of M.F. and E.S. is supported by the DOE under contract
DE-AC03-76SF00515, and by the A.P. Sloan Foundation, and that
of M.F. by a Stanford Graduate Fellowship.

\listrefs

\end